\documentclass[aps,prb,twocolumn,showpacs,preprintnumbers,amsmath,amssymb,superscriptaddress]{revtex4}
\usepackage{graphicx}
\usepackage{dcolumn}
\usepackage{bm}
\usepackage{longtable}
\usepackage{epsfig}
\usepackage{amssymb}
\usepackage{color}
\usepackage{subfigure,amsmath,verbatim,moreverb,bm}

\begin{document}
\title{Ground-state properties of the one-dimensional attractive Hubbard
model with confinement: a comparative study}
\author{Ji-Hong Hu}
\affiliation{Department of Physics, Zhejiang Normal University,
Jinhua 340012, China}
\author{Jing-Jing Wang}
\affiliation{Department of Physics, Zhejiang Normal University,
Jinhua 340012, China}
\author{Gao Xianlong}
\email{gaoxl@zjnu.edu.cn}
\affiliation{Department of Physics, Zhejiang Normal University,
Jinhua 340012, China}
\author{Masahiko Okumura}
\affiliation{CCSE, Japan Atomic Energy Agency, 6--9--3 Higashi-Ueno,
Taito-ku, Tokyo 110--0015, Japan} \affiliation{CREST (JST), 4--1--8
Honcho, Kawaguchi, Saitama 332--0012, Japan}
\author{Ryo Igarashi}
\affiliation{CCSE, Japan Atomic Energy Agency, 6--9--3 Higashi-Ueno,
Taito-ku, Tokyo 110--0015, Japan} \affiliation{CREST (JST), 4--1--8
Honcho, Kawaguchi, Saitama 332--0012, Japan}
\author{Susumu Yamada}
\affiliation{CCSE, Japan Atomic Energy Agency, 6--9--3 Higashi-Ueno,
Taito-ku, Tokyo 110--0015, Japan} \affiliation{CREST (JST), 4--1--8
Honcho, Kawaguchi, Saitama 332--0012, Japan}
\author{Masahiko Machida}
\affiliation{CCSE, Japan Atomic Energy Agency, 6--9--3 Higashi-Ueno,
Taito-ku, Tokyo 110--0015, Japan} \affiliation{CREST (JST), 4--1--8
Honcho, Kawaguchi, Saitama 332--0012, Japan}

\date{\today}
\begin{abstract}
 We revisit the one-dimensional attractive Hubbard model by using
 the Bethe-ansatz based density-functional theory and density-matrix
 renormalization method. The ground-state properties of this model are
 discussed in details for different fillings and different confining
 conditions in weak-to-intermediate coupling regime. We investigate the
 ground-state energy, energy gap, and pair-binding energy and compare
 them with those calculated from the canonical Bardeen-Cooper-Schrieffer approximation. We
 find that the Bethe-ansatz based density-functional theory
 is computationally easy and yields an accurate description of the
 ground-state properties for weak-to-intermediate interaction strength,
 different fillings, and confinements. In order to characterize the
 quantum phase transition in the presence of a harmonic confinement,
 we calculate the thermodynamic stiffness, the density-functional
 fidelity, and fidelity susceptibility, respectively. It is shown that
 with the increase of the number of particles or attractive interaction
 strength, the system can be driven from the Luther-Emery-type phase to
 the composite phase of Luther-Emery-like in the wings and
 insulating-like in the center.
\hspace*{7.0cm}
\end{abstract}
\maketitle
\section{introduction}
The Hubbard Hamiltonian is a simplified but important prototype
model for strongly correlated electrons in solid state. The
one-dimensional (1D) version is usually used as a basic model examining
the electronic correlation behavior in the quasi-one-dimensional systems
exhibiting the itinerant character of electrons and superconductivity
\cite{giamarchi,Essler} like the organic superconductor
(TMTSF)$_2$X.~\cite{Bourbonnais} Due to its solvability by the
Bethe-ansatz method,~\cite{LiebWu} this model serves as a popular benchmark
to test the accuracy of different approximate methods in handling the
ground-state (GS) properties of strongly correlated fermions especially
in an intermediate coupling
strength.~\cite{KocharianPRB,ShibaPRB,CollPRB} This model has recently
arisen renewed interest, in ultracold atomic gases, where fundamental
many-body phenomena can be investigated in a well-controlled way.

The ultracold atomic gases loaded on optical lattices have opened an
exciting field in simulating condensed-matter model Hamiltonians like
the boson and fermion Hubbard models. More recently a Tonks-Girardeau
gas of bosonic $^{87}$Rb atoms was realized experimentally in a 1D
optical lattice.~\cite{Paredes} In the case of fermionic atom gases,
however, it is more difficult to cool them down because of the Pauli
exclusion principle different from the bosonic atoms. Up to now, a
two-component Fermi gas of $^{40}$K atoms with tunable interacting
strengths has been prepared in a quasi-1D geometry.~\cite{moritz_2005}
An interacting Fermi gas of $^{40}$K atoms has been demonstrated in
three-dimensional optical lattices.~\cite{Kohl} Within present-day
techniques,~\cite{Pezze} it is possible to trap interacting gases of
fermionic atoms in 1D optical lattices and then simulate the 1D Hubbard
model experimentally. At the same time, experimental achievements on
population-imbalanced ultracold fermions \cite{Zwierlein,Shin} make it
possible to systematically study the exotic paring states in 1D fermion
systems with or without optical lattices.~\cite{unbalanced} Moreover,
the square-well trap and optical box-like trap have been experimentally
realized in the atomic system, which makes it meaningful to study the
physics of 1D uniform atomic gases with a hard wall
directly.~\cite{boxtrap}

Considering the strongly correlated nature of 1D systems and the
enhanced quantum fluctuations, it is crucial to have techniques beyond
the mean field theory, both analytically and numerically. Among them,
Bethe-ansatz solution,~\cite{LiebWu} quantum Monte-Carlo (QMC)
simulation,~\cite{Rigol,Rigol2004,Astrakhardik,Casula,Hirsch,Sandvik}
density-matrix renormalization group
(DMRG),~\cite{Machida1,gaoprl98,Molina,Schollwock} and some
sophisticated theoretical methods, such as the dynamical mean field
theory combined with the numerical renormalization group
\cite{Bauer0901} and a time-evolving block decimation numerical
study~\cite{wang0901}, are listed as well-established techniques. On the other hand, the exact diagonalization scheme
\cite{Machida1,Nikkarila} has also been used to obtain the exact
numerical solution for the small size systems as an ultimate technique.

The inhomogeneity caused by the boundary conditions or the external
potentials, normally invalidates a reliable analytical method that is
usually used in the homogeneous system. On the other hand, the numerical
schemes mentioned above, e.g., the exact Bethe-ansatz solution
incorporated with local density approximation is proved to be useful in
obtaining the phase diagram and the collective oscillations of the
atomic mass density in the presence of harmonic traps.~\cite{Hu} The
lattice version of the Bethe-ansatz based density-functional theory
(BADFT) \cite{Gunnarsson,Lima,gaoprl98,Schenk} is also a good candidate
in studying inhomogeneous strongly-correlated system. For the strongly interacting regime,
some non-perturbative methods, like, the
Fermi-Bose mapping and the Bethe-ansatz method, have to be
used.~\cite{LiebWu,Girardeau07,ChenShuPRL,Duan} In general, the
Bethe-ansatz solution provides reliable results beyond the mean-field
theory.~\cite{Batchlor} BADFT has been used in studying the 1D
Gaudin-Yang gases \cite{Magyar,Gaopra73} and 1D Hubbard
model,~\cite{Lima,gaoprb73} which is different from the density-matrix
functional theory with the basic variable to be the single-particle
density matrix.~\cite{Sandoval}

The rich non-uniform phases induced by the trapping potential can be
identified by the variance of the local density, the local
compressibility,~\cite{Rigol2004} the double occupancy,~\cite{Kollath}
or the thermodynamic stiffness.~\cite{gaoprb73rapid} The entanglement
entropy is found by Fran\c{c}a et al. to be a powerful measure
characterizing the phase-separated states in spatially inhomogeneous
environment.~\cite{entanglement} Recently, a new concept named fidelity
in quantum-information and quantum-computation theory, is used to
characterize quantum phase transitions by computing the overlap between
two states of nearby points in parameter space. By measuring the
similarity between these two states, it is expected to show a minimum in
the fidelity due to a dramatic change in the ground-state wavefunctions
around the critical point of the quantum phase transition. The studies
are made both experimentally and theoretically on many model systems,
such as on the 1D XY model, the Dicke model, and the 1D Hubbard
model.~\cite{fidelity} Gu further proposed a density-functional fidelity
to measure the similarity between density distributions of two ground
states in parameter space based on the observation that the density
distributions \cite{fidelity_Gu} can reflect the change in the
ground-state wave functions. An interesting question then arises:
Can this method be applied to quantify the composite quantum phases
transitions of the 1D Hubbard model in the presence of confinements?

The 1D Fermi-Hubbard model with attractive on-site interactions, has
long served a ``minimal'' model that best describes
superconductivity. For the fermionic atomic gases loaded on optical
lattices, the strength or even the sign of the on-site interaction
between different species can be easily tuned by using a technique named
Feshbach resonance.~\cite{Feshbach} Marsiglio calculated the
ground-state energy (GSE) and energy gap to the first excited state for
the attractive Hubbard model using the variational canonical
Bardeen-Cooper-Schrieffer (CBCS) and grand CBCS wave
function.~\cite{marsiglio_prb1997} This model for different
concentrations of electrons is also studied within a self-consistent
field method.~\cite{KocharianPRB}

In this work, motivated by earlier theoretical work and by ongoing
experimental efforts, we revisit the 1D attractive Hubbard model in the
weak-to-intermediate coupling strength under different boundary
conditions with or without trapping potential. We present a fully
microscopic theoretical study on the ground-state properties.

We begin with introducing the Bethe-ansatz solution of the model and
solving the coupled equations. In Section \ref{ldft}, we provide a
Bethe-ansatz based spin-density-functional theory (SDFT) and its
local-spin-density approximation. In Section \ref{results}, we show our
main numerical results, compared to the exact Bethe-ansatz, the CBCS
\cite{Tanaka_prb1999}, and DMRG ones. Our comparisons are done on general
band fillings and a wide range of coupling strength in three different
systems, i.e., 1D lattices with periodic boundary condition, hard-wall,
and harmonic confinement. The energy, the binding gap, and the
pair-binding energy of the ground state are in a good agreement with the
exact results. We also provide a simple and direct diagnostic of the
quantum phase transition by calculating the thermodynamic stiffness,
density-functional fidelity, and fidelity susceptibility. At last, our
results are summarized.

\section{The 1D attractive Fermi-Hubbard model}\label{sect:model}

We consider a 1D attractively interacting two-component Fermi gas with
$N_{\rm f}$ atoms in a lattice with unit lattice constant and $N_{\rm
s}$ lattice sites, which can be described by a one-band Fermi-Hubbard
model,~\cite{jaksch_98}
\begin{eqnarray}\label{eq:hubbard}
\hat{H}_{\rm s} & = & \hat {H}_{\rm ref}+\hat{H}_{\rm ext} \nonumber \\
& = & -t\sum_{i=1,\sigma}^{N_{\rm s}-1}({\hat
 c}^{\dag}_{i\sigma}{\hat c}^{\phantom \dag}_{i+1\sigma}+{\rm H.c.}) + U
 \sum_{i=1}^{N_{\rm s}}\,{\hat n}_{i\uparrow}{\hat n}_{i\downarrow}
 \nonumber \\
&& {} +\sum_{i=1,\sigma}^{N_{\rm s}} V_{i\sigma}{\hat n}_{i\sigma} \,
 ,
\end{eqnarray}
where $\sigma=\uparrow,\downarrow$ is a pseudospin-$1/2$ label for
two internal hyperfine states, ${\hat n}_i= \sum_{\sigma} {\hat
n}_{i\sigma}=\sum_{\sigma} {\hat c}^{\dag}_{i\sigma}{\hat c}^{\phantom
\dag}_{i\sigma}$ is the total site occupation operator, $t$ is
the tunneling between nearest neighbors, $U$ is the strength of the
on-site attractive interaction, and $V_{i\sigma}$ describes the trapping
potential. We consider in this paper 1D lattices with periodic
boundary condition, in a simple box-shaped trapping potential
($V_{i\sigma}=0$ inside the box and $V_{i\sigma}=+\infty$ elsewhere,
namely, a hard-wall) and a parabolic confinement ($V_{i\sigma}=V_2 [ i -
(N_{\rm s}-1)/2]^2$ with $V_2$ the amplitude of the confining
potential). The numbers of atoms with spin up and spin down are
$N_\uparrow$ and $N_\downarrow$, respectively.

The 1D attractive Fermi-Hubbard model in the absence of confining
potentials (i.e., $\hat{H}_{\rm s}=\hat{H}_{\rm ref}$) belongs to the
universality class of Luther-Emery liquids, characterized by a massive
spin gap and zero charge gap. At zero temperature, the properties of
$\hat{H}_{\rm ref}$ in the thermodynamic limit ($N_{\sigma}, N_{\rm s}
\rightarrow \infty$ with finite $N_{\sigma} / N_{\rm s}$) are determined
by the filling $n_\sigma = N_\sigma / N_{\rm s}$ and by the
dimensionless coupling constant $u=U/t$. According to Lieb and
Wu~\cite{LiebWu}, the ground state of $\hat {H}_{\rm ref}$ for different
fillings in the thermodynamic limit is described by the coupled integral
equations for the momentum distributions $\rho(k)$ and
$\sigma(\lambda)$,
\begin{eqnarray}
\rho(k) &= &\frac{1}{2\pi}+\frac{|u|\cos(k)}{4\pi}
\int_{-B}^{B}d\lambda \, f_1(k,\lambda)\sigma(\lambda) \,, \\
\sigma(\lambda)&=&\frac{|u|}{4\pi} \int_{-Q}^{Q} dk \,
 f_1(k,\lambda)\rho(k)\nonumber \\
&& {} - \frac{|u|}{2\pi} \int_{-B}^{B} d \lambda' \, f_2
 (\lambda,\lambda') \sigma(\lambda') \,,
\end{eqnarray}
with
\begin{eqnarray}
f_1(k,\lambda) & = & \frac{1}{(u/4)^2+(\lambda-\sin k)^2} \,, \\
f_2(\lambda,\lambda') & = &
 \frac{1}{(u/2)^2+(\lambda-\lambda')^2} \,.
\end{eqnarray}
The non-negative parameters $Q$ and $B$ are determined by the
normalization conditions $\int_{-Q}^{Q} dk \, \rho(k)=1-2s$ and
$\int_{-B}^{B} d\lambda \, \sigma(\lambda) = (n-2s)/2$, where the
magnetization $s$ is defined as $s = (N_{\uparrow}-N_{\downarrow}) /
(2N_{\rm s})$. The ground-state energy per site of the system is written
in terms of the momentum distributions as,
\begin{eqnarray}\label{eq:GSE}
\epsilon_{\scriptscriptstyle \rm GS}(n,s,u) = - |U|
 \left[\frac{n}{2}-s\right] -2t\int_{-Q}^{Q}d k \, \rho(k)\cos k\,,
\end{eqnarray}
which is a function of filling $n$, magnetization $s$, and interaction
strength $u$. The GS properties of the 1D attractive Hubbard model have
been analyzed by many authors.~\cite{KocharianPRB,AHM}

\section{Lattice spin-density-functional theory}\label{ldft}

Density functional theory (DFT) \cite{dft,Giuliani_and_Vignale} is a
powerful tool to calculate the GS properties of a many-body Hamiltonian
including inhomogeneity. Through Kohn-Sham mapping, the true system of
interacting particles is mapped onto a fictitious one in which the
particles do not interact but reproduce the true particle
density. The central problem in DFT is the exchange-correlation (xc)
functional, which remains elusive and can be simply treated by the
so-called local-density approximation (LDA).~\cite{Kohn} The essence of
LDA is to locally approximate the xc energy of the interacting
inhomogeneous system based on that of an interacting homogeneous
reference system.
\begin{figure}
\centering
\includegraphics*[width=1.00\linewidth]{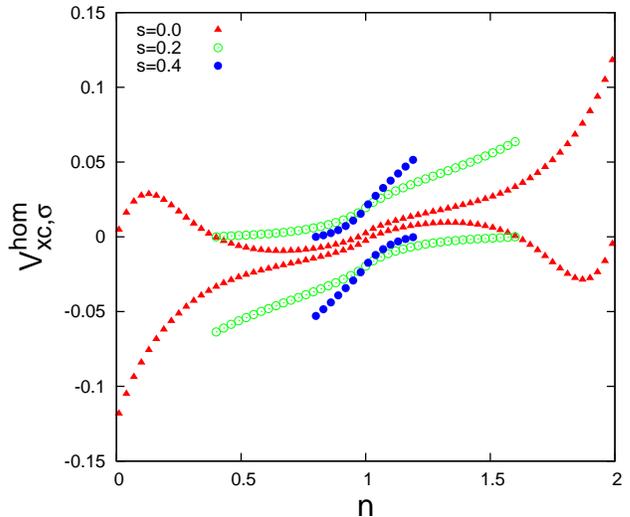}
\includegraphics*[width=1.00\linewidth]{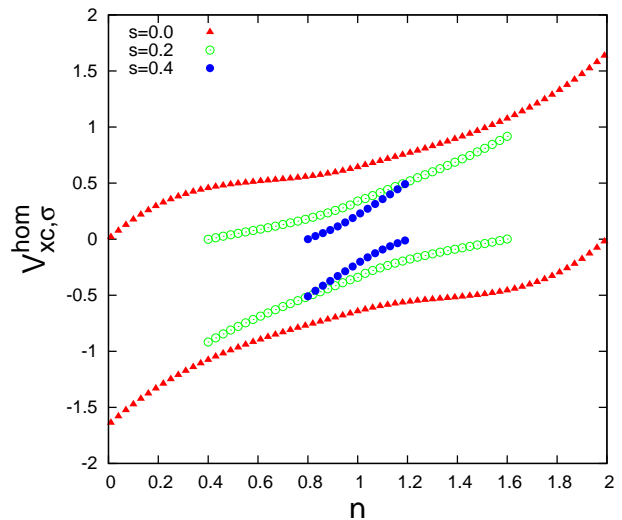}
\caption{(Color online) Top panel: The exchange-correlation potential (in units of $t$) of
 the 1D attractive Hubbard model as a function of $n$
 for various values of $s$ with $u=-1$. The upper three data are for
 $V^{\rm hom}_{\rm xc,\uparrow}$ and the lower three for $V^{\rm
 hom}_{\rm xc,\downarrow}$. Obviously there is no charge gap for the
 attractive Hubbard model. Bottom panel: The same as that in the top
 panel but for $u=-4$.\label{fig:one}}
\end{figure}
\begin{figure}
\begin{center}
\includegraphics*[width=1.00\linewidth]{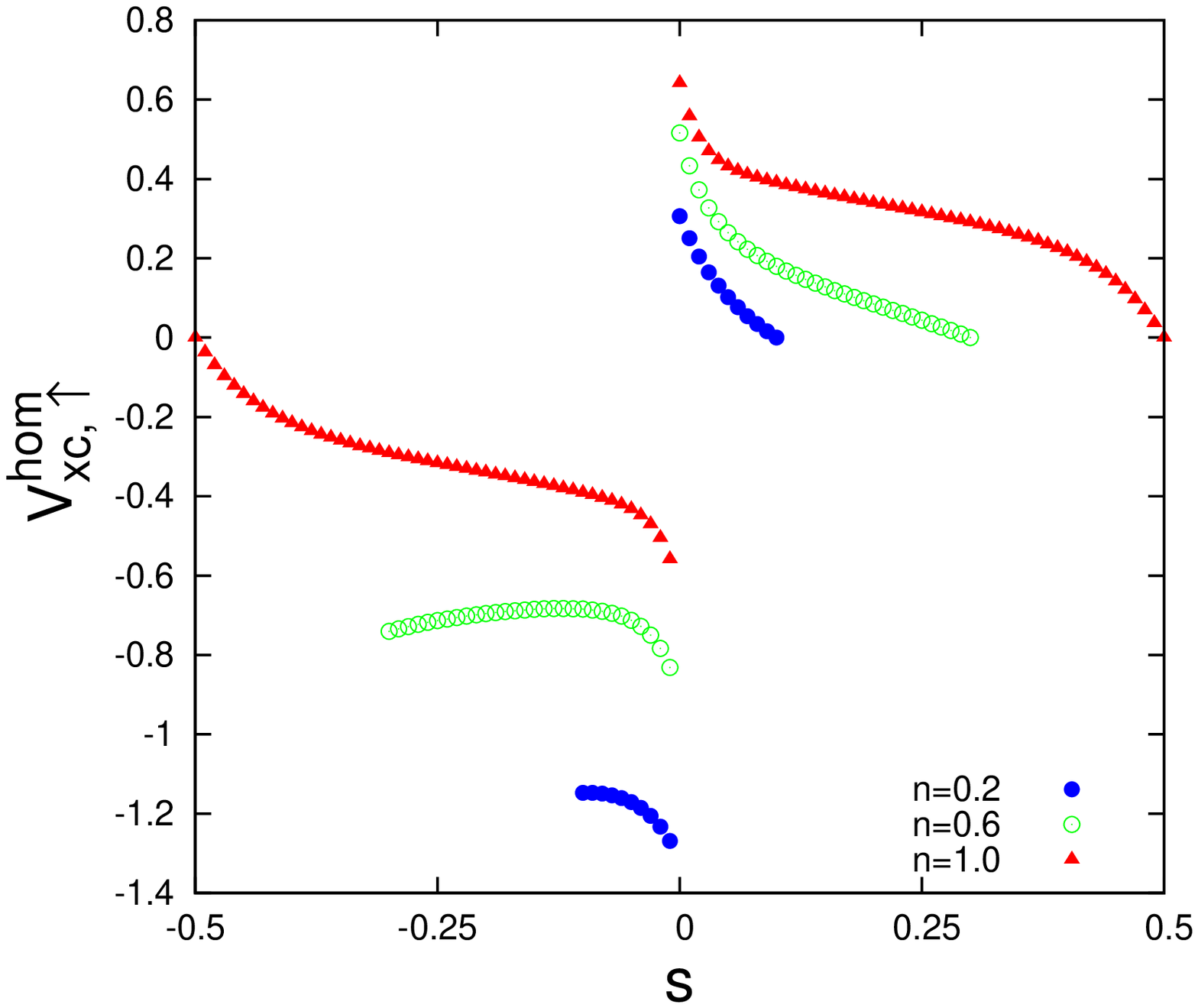}
\includegraphics*[width=1.00\linewidth]{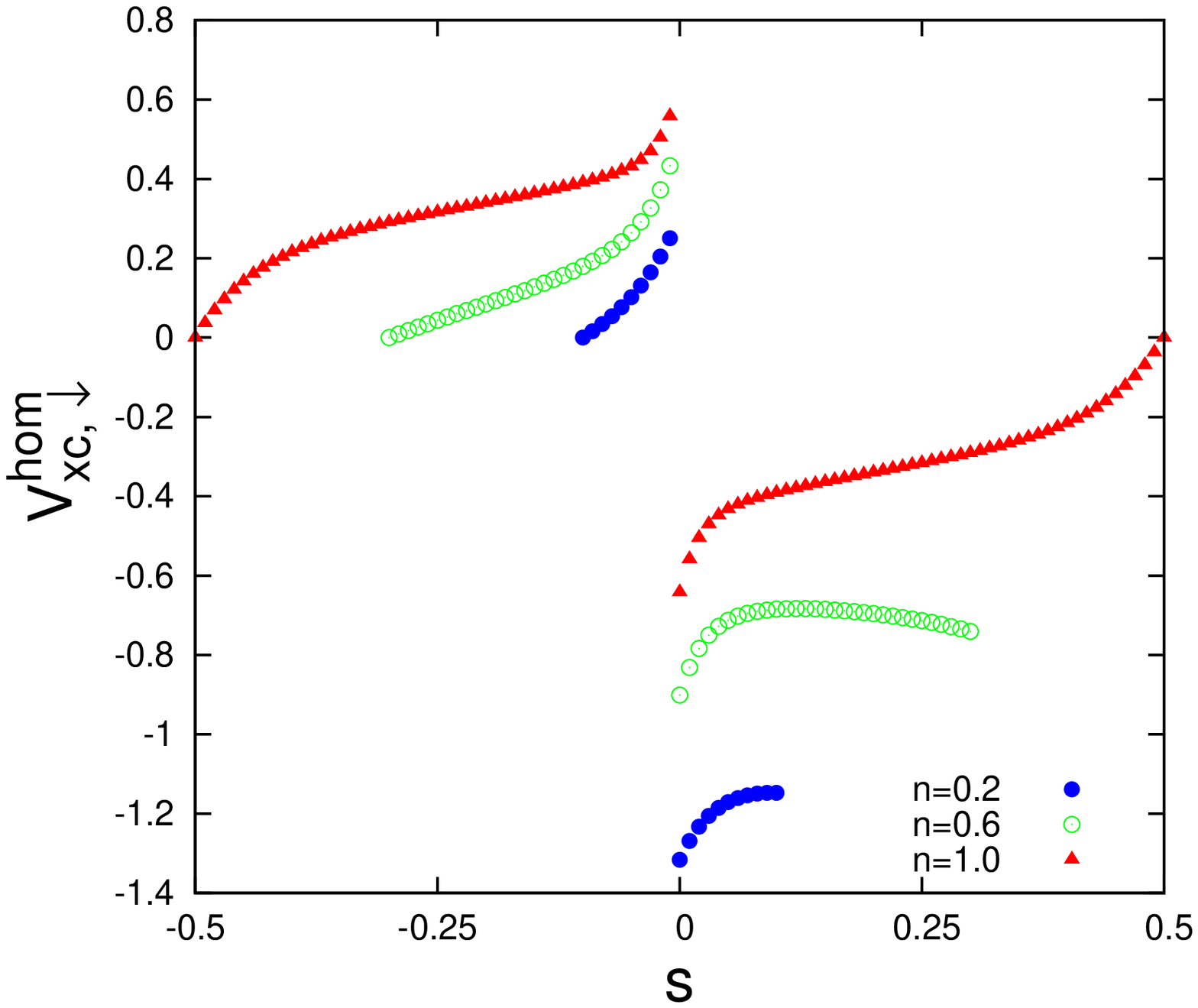}
\caption{(Color online) Top panel: The exchange-correlation potential (in units
 of $t$) $V_{\rm xc, \uparrow}^{hom}$ of the attractive Hubbard model as a function of $s$ for various values of $n$ with $u=-4$. The
 spin channel opens a gap at $s=0$.  Bottom panel: The same as that in
 the top panel but for $V_{\rm xc, \downarrow}^{hom}$.\label{fig:two}}
\end{center}
\end{figure}
In discrete lattice systems, there is a lattice-version DFT, the
so-called site-occupation functional theory (SOFT), introduced in the
pioneering papers by Gunnarsson and Sch\"{o}nhammer to study the
band-gap problem in the context of {\it ab initio} theories of
fundamental energy gaps in semiconductors and
insulators.~\cite{Gunnarsson} Within SOFT the GS site occupation can be
obtained by solving self-consistently the lattice Kohn-Sham (KS)
equations
\begin{equation}\label{eq:sks}
\sum_{j=1}^{N_{\rm s}} \left( - t_{i,j} + V^{\rm \scriptscriptstyle
KS}_{i\sigma}[n_{i\sigma}] \delta_{ij} \right)
\varphi^{(\alpha)}_{j\sigma} = \epsilon^{(\alpha)}_\sigma
\varphi^{(\alpha)}_{i\sigma}
\end{equation}
together with the closure relation
\begin{equation}\label{eq:closure}
n_{i\sigma} = \sum_{\alpha=1}^{N_\sigma} \left| \varphi^{(\alpha)}_{i\sigma}
\right|^2 \,.
\end{equation}
Here the sum $\alpha$ runs over all the occupied orbitals, and the
effective KS potential is given by $V^{\rm \scriptscriptstyle
KS}_{i\sigma}[n_{i\sigma}] =U n_{i\bar{\sigma}}+V^{\rm
xc}_{i\sigma}[n_{i\sigma}]+V^{\rm ext}_{i\sigma}$ where
$\bar{\sigma}=-\sigma$. The first term in the Kohn-Sham potential comes
from the Hartree mean-field contribution, while $V^{\rm
\scriptstyle xc}_{i\sigma}[n_{i\sigma}]$ is the derivative of the
xc energy $E_{\rm xc}[n_\uparrow, n_\downarrow]$ evaluated at the GS
site occupation.

The total GSE of the system is given by
\begin{eqnarray}\label{eq:gs_energy}
E_0 [ n_\uparrow, n_\downarrow ] & = & \sum_{\sigma} \sum_\alpha
 \epsilon^{(\alpha)}_\sigma - \sum_\sigma \sum_i V^{\rm xc}_{i\sigma}
 n_{i\sigma} \nonumber \\
&& {} - \sum_\sigma \sum_i U n_{i\sigma} n_{i\bar{\sigma}} + E_{\rm xc}
 [n_\uparrow, n_\downarrow] .
\end{eqnarray}
The local-spin-density approximation (LSDA) for the spin-polarized
system has been shown to provide an excellent account of the GS
properties over a large variety of inhomogeneous systems. In this work
we employ the following Bethe-ansatz based LSDA functional (BALSDA),
\begin{equation}\label{eq:balda}
\left. V^{\rm xc}_{i\sigma} \right|_{\rm BALSDA} = \left. V^{\rm
 hom}_{\rm xc,\sigma} (n,s,u) \right|_{n\rightarrow n_i, s\rightarrow
s_i} ,
\end{equation}
where the xc potential of the 1D homogeneous Hubbard model is defined
by~\cite{gaoprb73}
\begin{equation}\label{eq:vxchom}
V^{\rm hom}_{\rm xc,\sigma} (n,s,u) = \frac{\partial}{\partial n_\sigma}
\left[ \epsilon_{\scriptscriptstyle \rm GS} (n,s,u) -
\epsilon_{\scriptscriptstyle \rm GS} (n,s,0) - U n_{i\sigma}
n_{i\bar{\sigma}} \right] . \nonumber
\end{equation}
Here, $\epsilon_{\scriptscriptstyle \rm GS}(n,s,u)$ is the GS energy per
site of the 1D homogeneous system defined in Eq.~(\ref{eq:GSE}) and
$\epsilon_{\scriptscriptstyle \rm GS}(n,s,0) = - (4t/\pi) \sin (
\pi n/2) \cos(\pi s)$.
\begin{figure}
\begin{center}
\includegraphics*[width=1.0\linewidth]{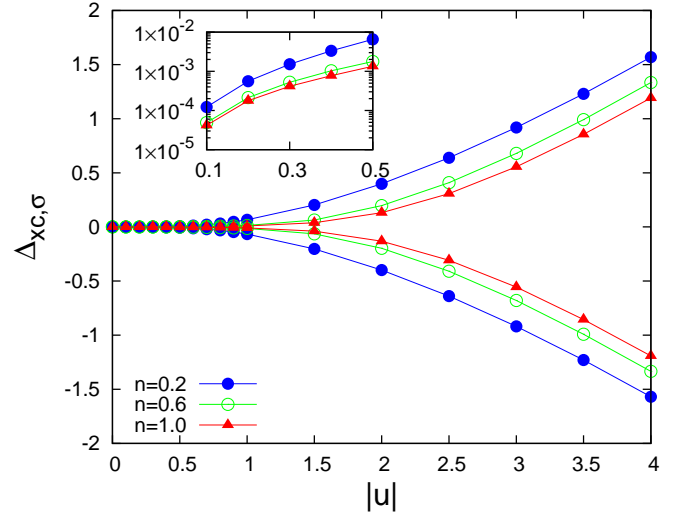}
\caption{(Color online) The xc gap $\Delta_{\rm xc,\sigma}(n,u)$ (in
 units of $t$) as a function of $|u|$. In the positive direction,
 $\Delta_{\rm xc,\uparrow}(n,u)$ are shown and in the negative direction
 $\Delta_{\rm xc,\downarrow}(n,u)$ are plotted. In the inset, the semilog plot for
 $\Delta_{\rm xc,\uparrow}(n,u)$ is shown for $|u|=0.1$ to $0.5$.
 \label{fig:three}}
\end{center}
\end{figure}
\begin{figure}
\begin{center}
\includegraphics*[width=0.9\linewidth]{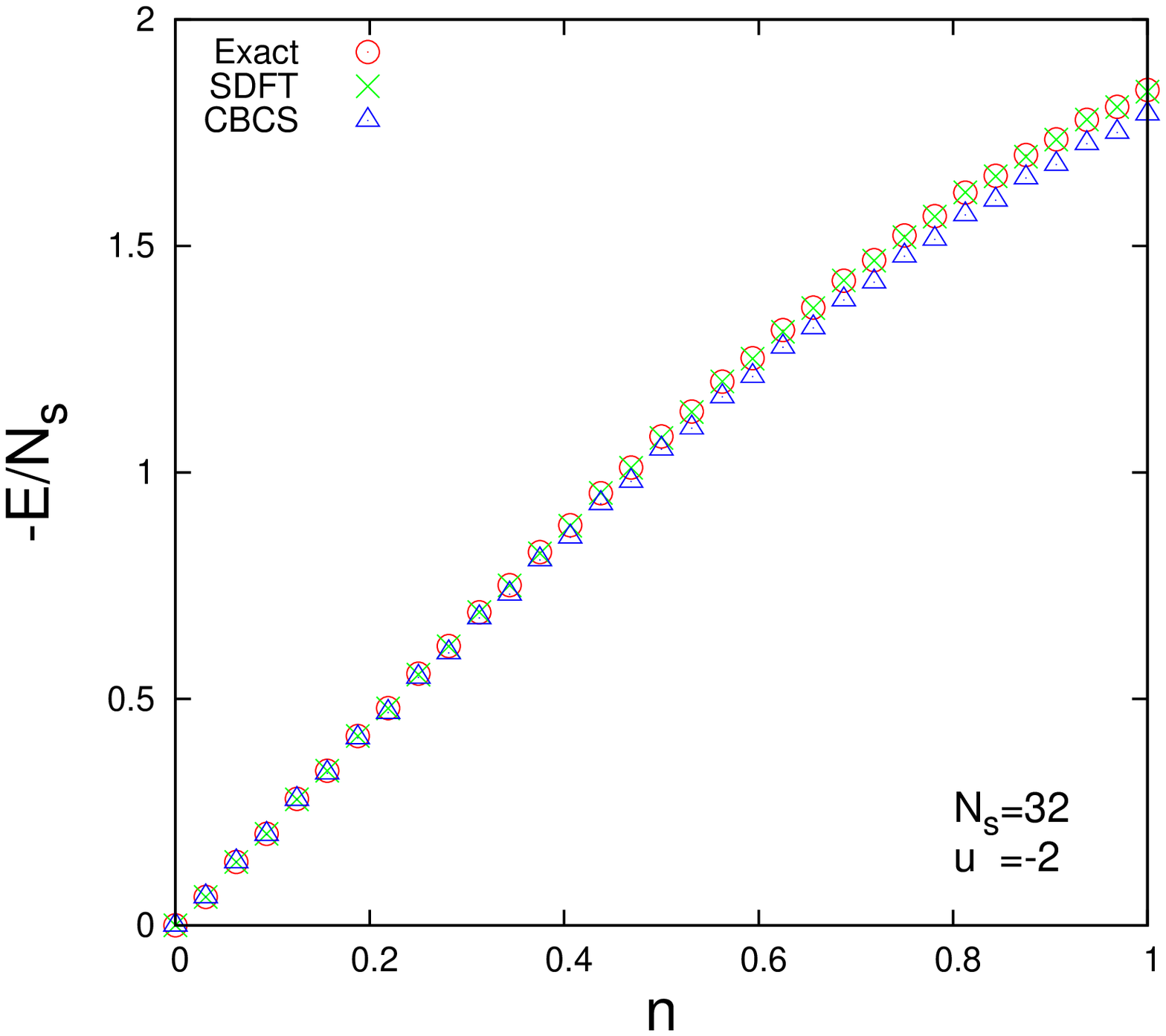}
\includegraphics*[width=0.9\linewidth]{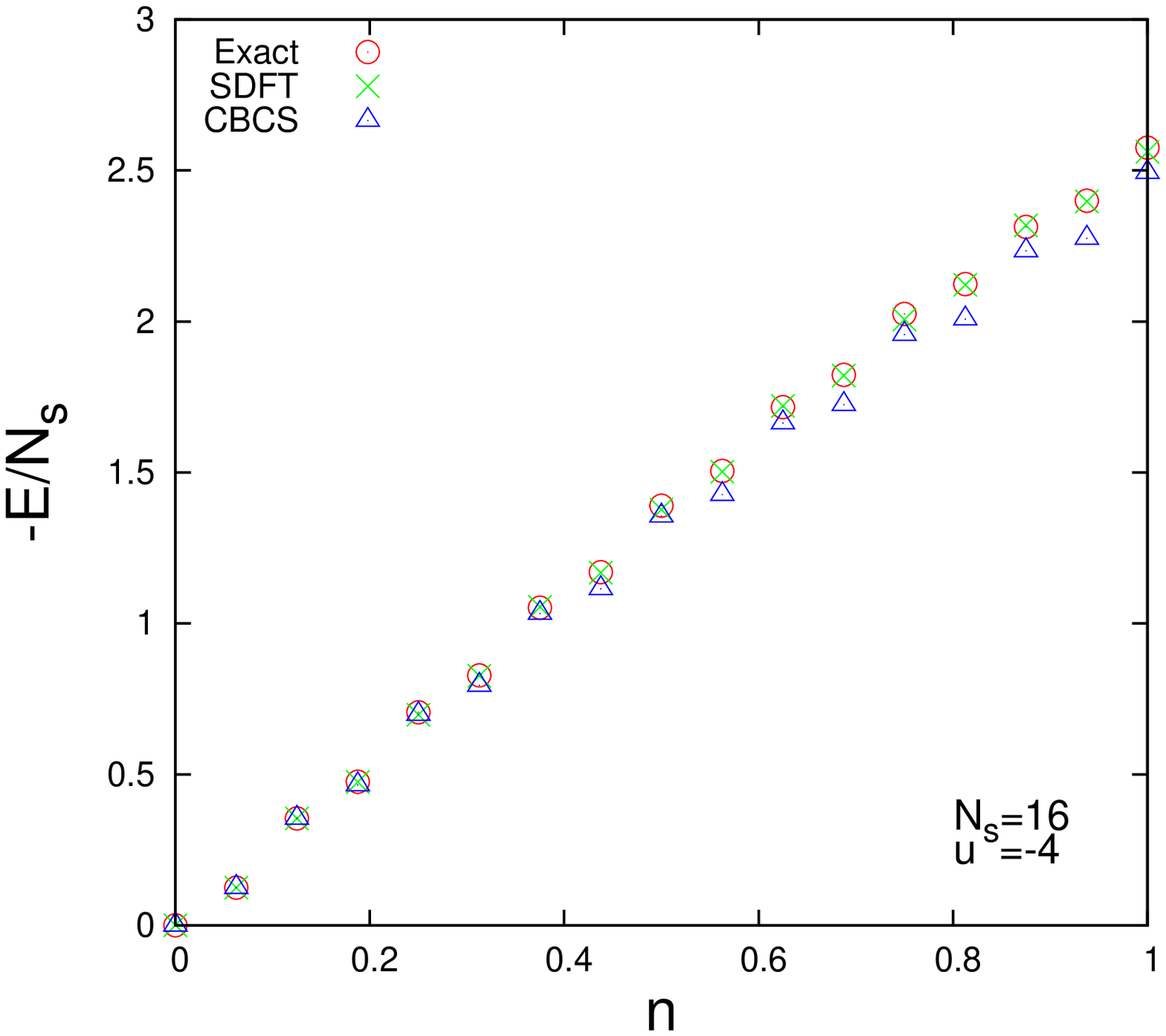}
\caption{(Color online) Top panel: The ground-state energy per site (in units of $t$) of
 the attractive 1D Hubbard model with periodic
 boundary condition as a function of particle density $n = N_{\rm f} /
 N_{\rm s}$ for $u=-2$ in a lattice with $N_{\rm s} = 32$
 sites. SDFT/BALSDA results (crosses) are compared with the CBCS data
 (triangles) and the exact results (open circles). The CBCS data are
 taken from Ref.~\onlinecite{Tanaka_prb1999} and the exact data are
 calculated from the couple Bethe-ansatz equations. Bottom panel: The
 same as that in the top panel but for $u=-4$ in a lattice with $N_{\rm
 s} = 16$ sites. The system of even number of particles is implicit with zero magnetization ($s=0$).
 For the system of odd number of particles, we choose $N_\uparrow-N_\downarrow=1$, that is, $s=1/(2N_s)$.
 In this case, SDFT has to be used to obtain an accurate GS energy.
 \label{fig:four}}
\end{center}
\end{figure}

Making use of the chain rule,
\[
\frac{\partial}{\partial n_\sigma} = \frac{\partial n}{\partial
n_\sigma} \frac{\partial}{\partial n} + \frac{\partial s}{\partial
n_\sigma} \frac{\partial}{\partial s},
\]
we obtain an explicit expression for the exchange-correlation potential
in terms of the chemical potential $\mu(n,s,u)$ and magnetic field
$h(n,s,u)$,
\begin{eqnarray}
V^{\rm hom}_{\rm xc,\sigma} (n,s,u) & = & \mu(n,s,u) \pm \frac{1}{2}
 h(n,s,u) \nonumber\\
&&{} + 2 t \cos \left[ \pi \left( \frac{n}{2} \pm s \right) \right] - U
 n_{\bar{\sigma}} \, ,
\end{eqnarray}
where the upper (lower) sign refers to the spin-up (spin-down) atomic
species. The chemical potential and magnetic field are defined by the
first derivative of the GSE of the homogeneous system with respect to
filling $n$ and magnetization $s$ as,
\begin{eqnarray}
\mu (n,s,u) & = & \frac{\partial}{\partial n}
\epsilon_{\scriptscriptstyle\rm GS}(n,s,u) \, \\
h (n,s,u) & = & \frac{\partial}{\partial s}
\epsilon_{\scriptscriptstyle\rm GS} (n,s,u) \, .
\end{eqnarray}

In Fig.~\ref{fig:one}, we illustrate the numerical results for the xc
potential as a function of filling $n$ for different magnetization
$s$. Obviously, the xc potential is continuous in the whole range of $n$
and there is no charge gap for the system of attractive interactions,
which is different from the repulsive case where a discontinuity occurs
and a charge gap opens at $n=1$.~\cite{gaoprb73}

In Fig.~\ref{fig:two}, we plot the xc potential as a function of $s$ for
various values of $n$ with $u=-4$. We see that there is always a
discontinuity at $s=0$, which corresponds to a gap opened in the spin
channel signalling a Luther-Emery liquid phase of paired particles. We
define the xc gap $\Delta_{\rm xc,\sigma}$ as the discontinuity in
$V^{\rm hom}_{\rm xc,\sigma}(n,s,u)$ at $s=0$,
\begin{eqnarray}\label{eq:delta_xc}
\Delta_{\rm xc,\sigma} (n,u) = \lim_{s\rightarrow 0^+} V^{\rm hom}_{\rm
 xc,\sigma}(n,s,u) - \lim_{s\rightarrow 0^-} V^{\rm hom}_{\rm xc,\sigma}
 (n,s,u) \,. \nonumber
\end{eqnarray}

The results are shown in Fig.~\ref{fig:three} for different fillings. We
notice there is a symmetry between $\Delta_{\rm xc,\uparrow}(n,u)$ and
$\Delta_{\rm xc,\downarrow}(n,u)$, as a result of the symmetries of the
GSE $\epsilon_{\scriptscriptstyle \rm GS} (n,s,u) =
\epsilon_{\scriptscriptstyle \rm GS}(n,-s,u)$ and our choice for the
Hartree energy. As a consequence, the xc potential in
Eq.~(\ref{eq:balda}) possesses a discontinuity at $s=0$ and naturally
the information of the Luther-Emery nature and the spin energy gap in
the reference system is transferred to the inhomogeneous system through
Eq.~(\ref{eq:balda}).

\section{Numerical results}\label{results}
\begin{figure*}
\begin{center}
\tabcolsep=0 cm
\begin{tabular}{cc}
\scalebox{0.5}[0.5]{\includegraphics{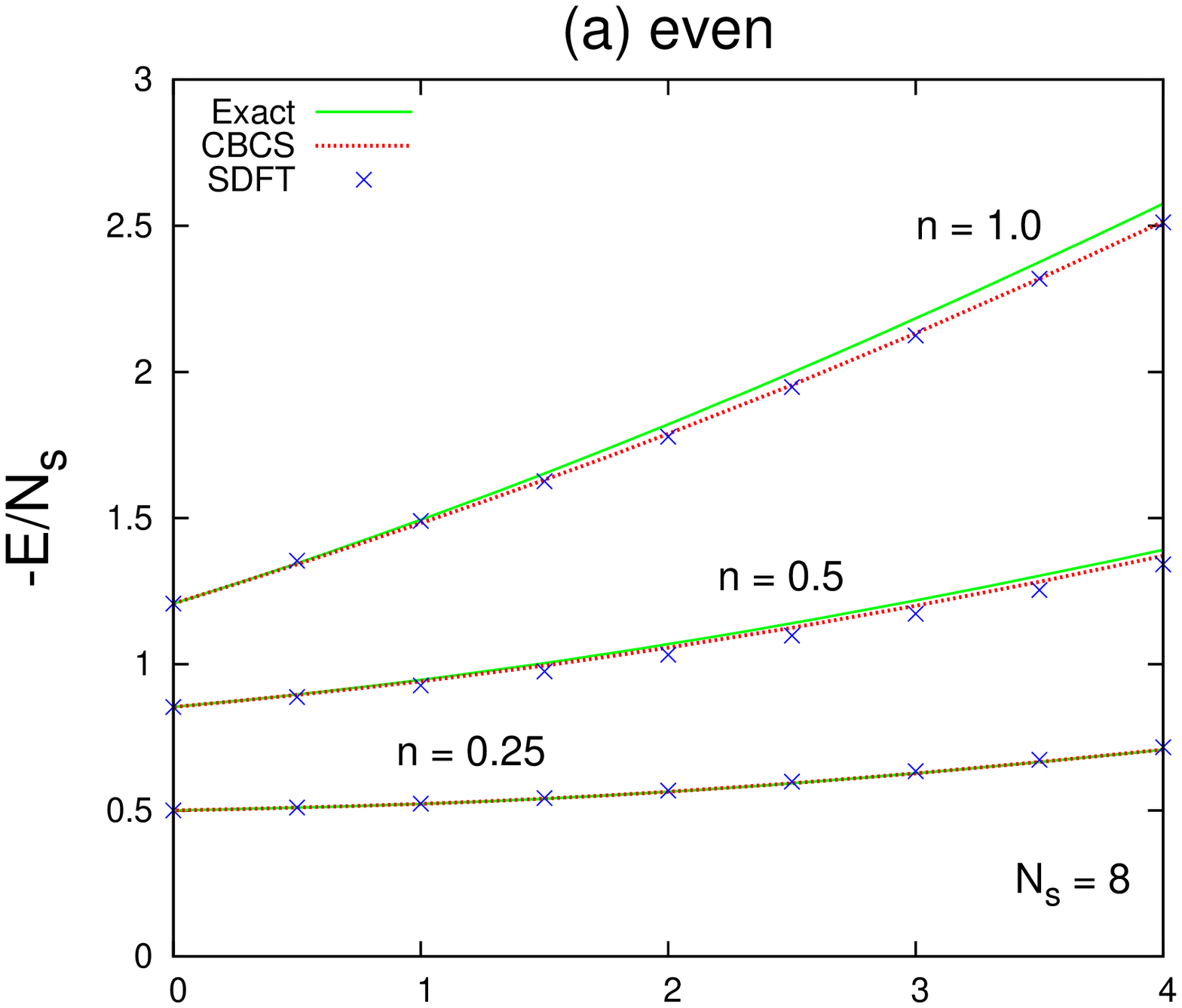}}&
\scalebox{0.5}[0.5]{\includegraphics{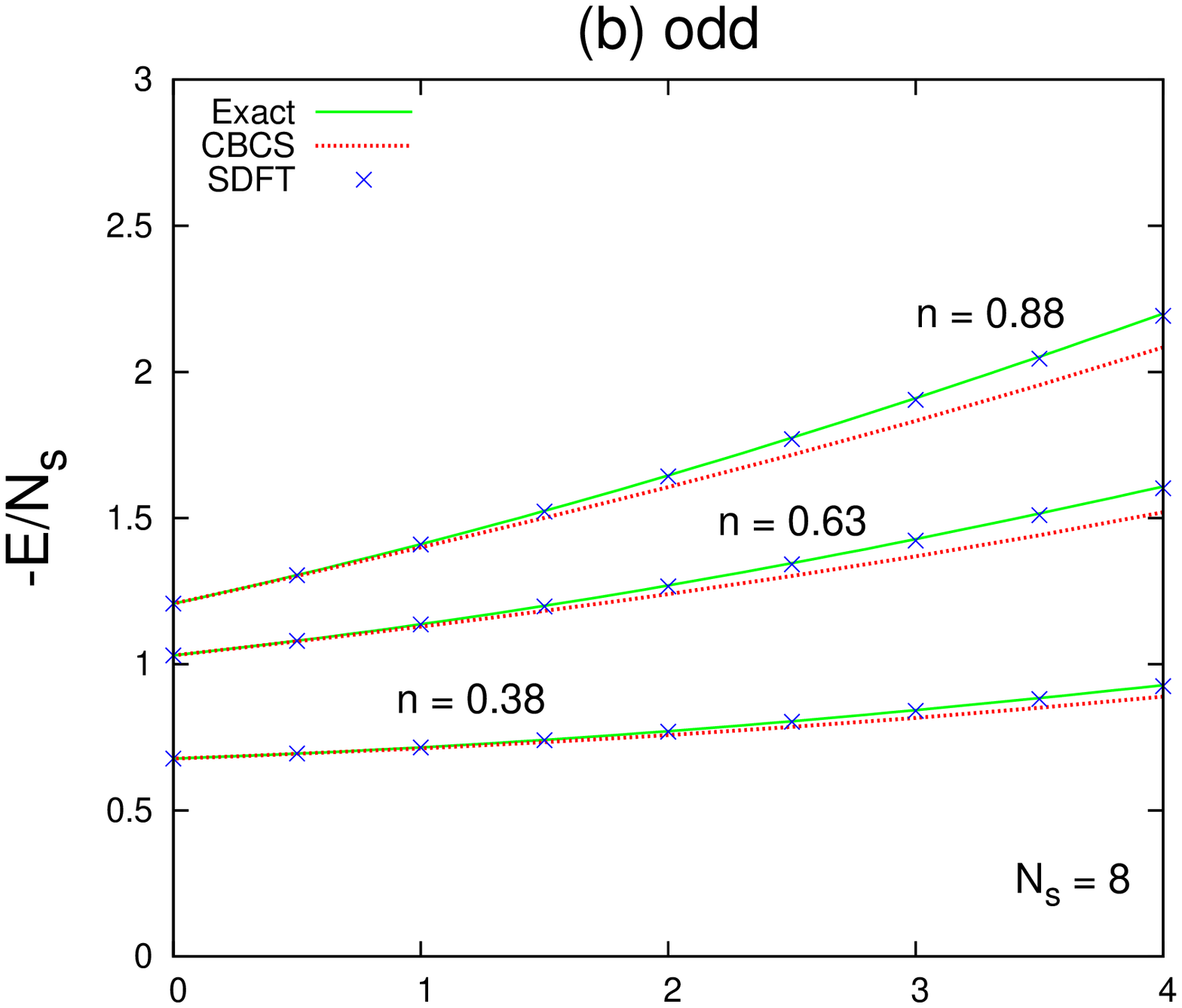}}\\
\scalebox{0.5}[0.5]{\includegraphics{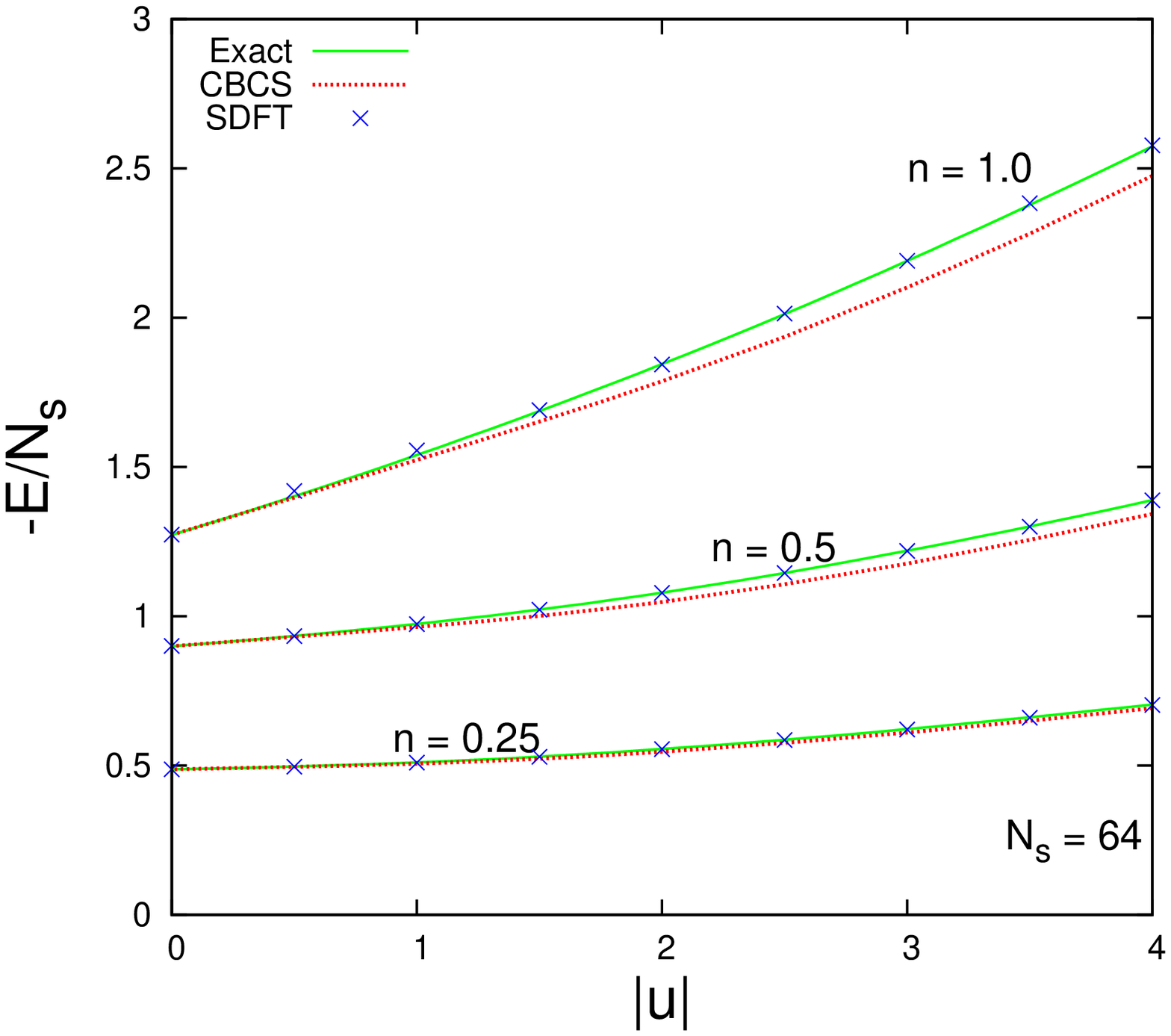}}&
\scalebox{0.5}[0.5]{\includegraphics{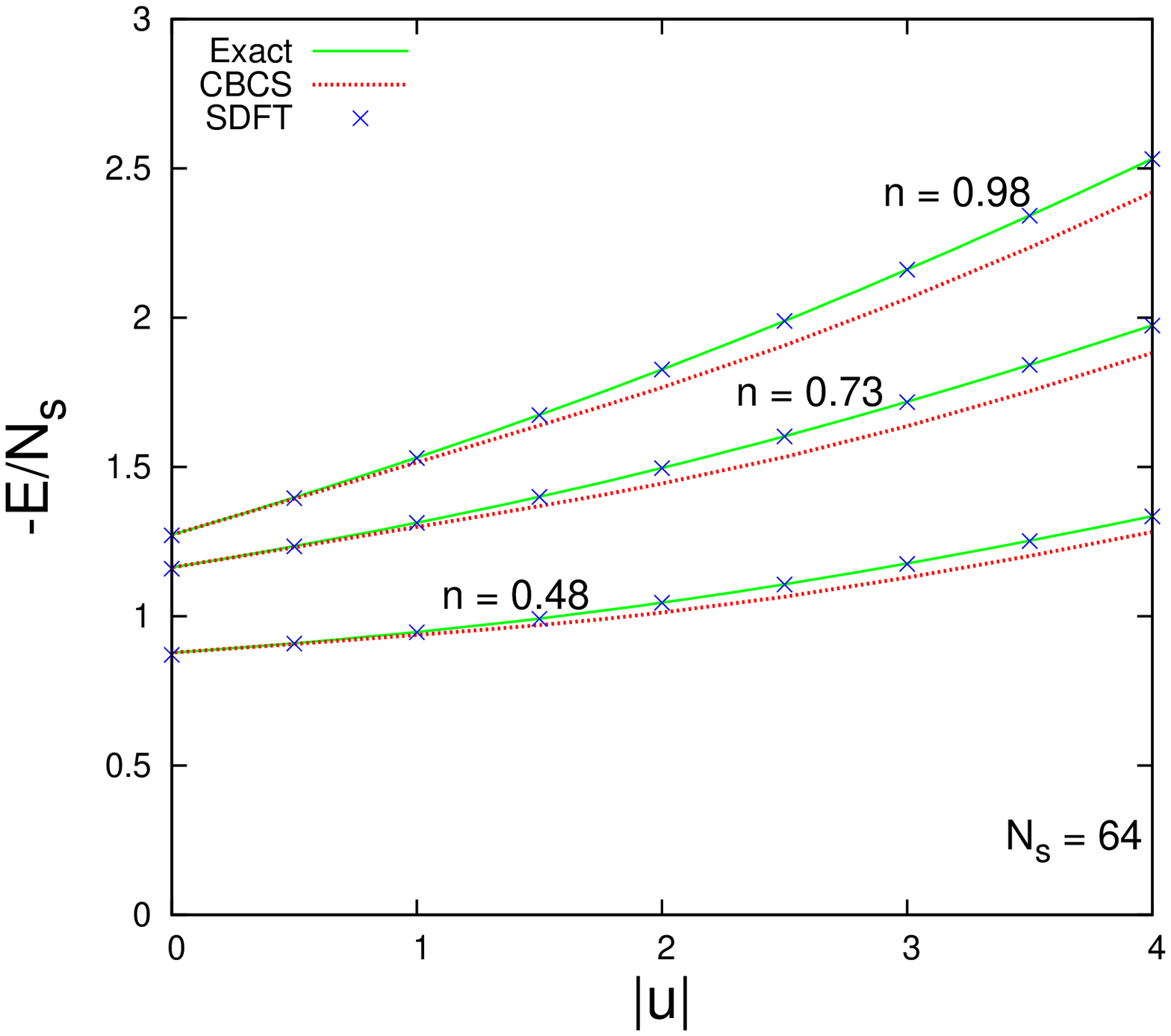}}
\end{tabular}
\caption{(Color online) The ground-state energy per site (in units of $t$) of the
 attractive 1D Hubbard model with periodic boundary
 condition as a function of the coupling strength $u$ with even (left
 panel) and odd (right panel) number of particles in a lattice of
 $N_{\rm s} = 8$ (upper frame) and $N_{\rm s} = 64$ (lower frame) sites
 for various values of fillings. SDFT/BALSDA results ($\times$) are
 compared with the exact (solid line) and CBCS (dotted lines) ones. (a)
 For $N_{\rm s} = 8$ and $N_{\rm s} = 64$ with even numbers of
 particles (the magnetization $s=0$). (b) The same as (a), but with odd numbers of
 particles (the magnetization $s=1/(2N_s)$, and $N_\uparrow-N_\downarrow=1$). \label{fig:five}}
\end{center}
\end{figure*}

\begin{figure*}
\begin{center}
\tabcolsep=0 cm
\begin{tabular}{cc}
\scalebox{0.5}[0.5]{\includegraphics{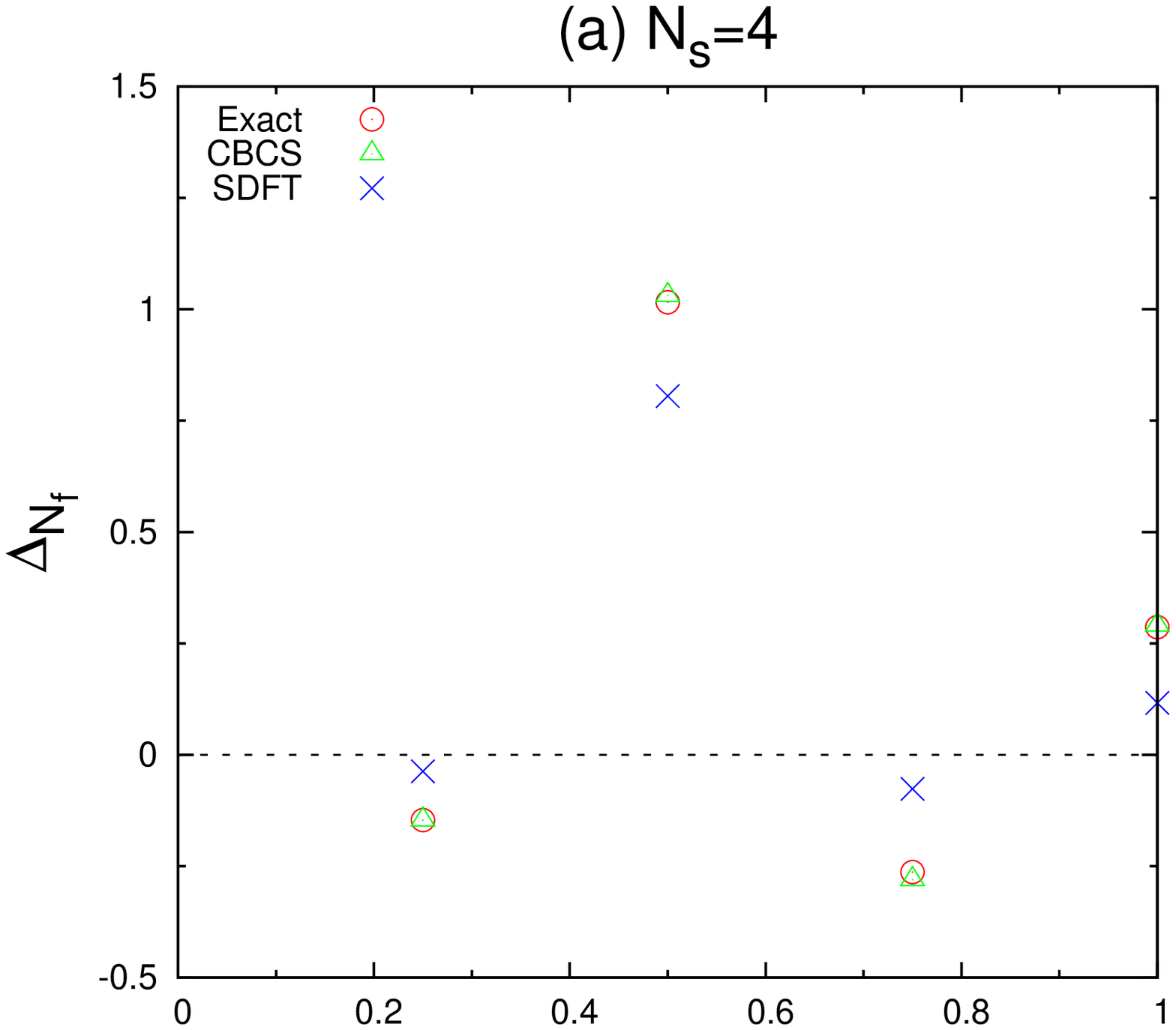}}&
\scalebox{0.5}[0.5]{\includegraphics{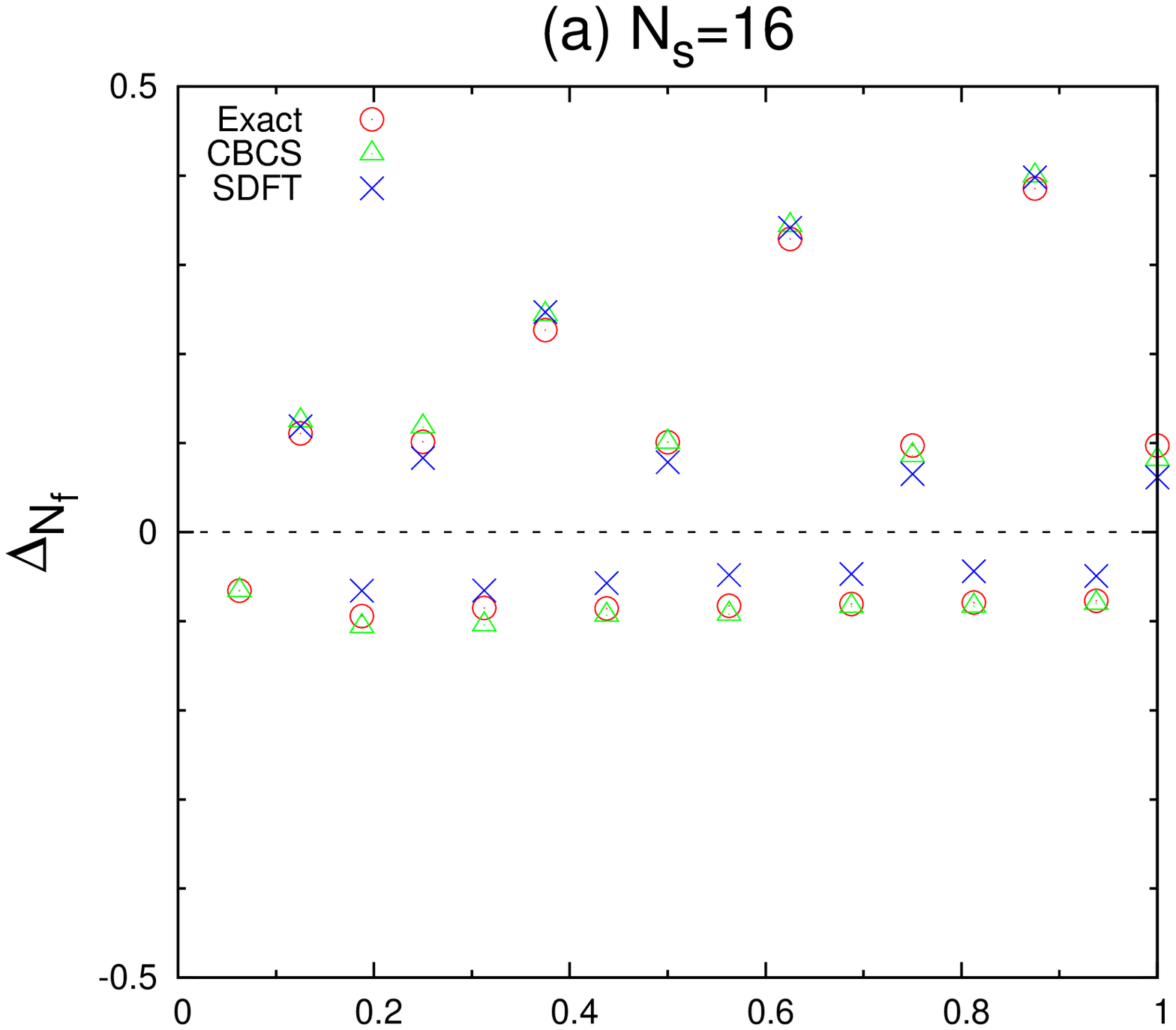}}\\
\scalebox{0.5}[0.5]{\includegraphics{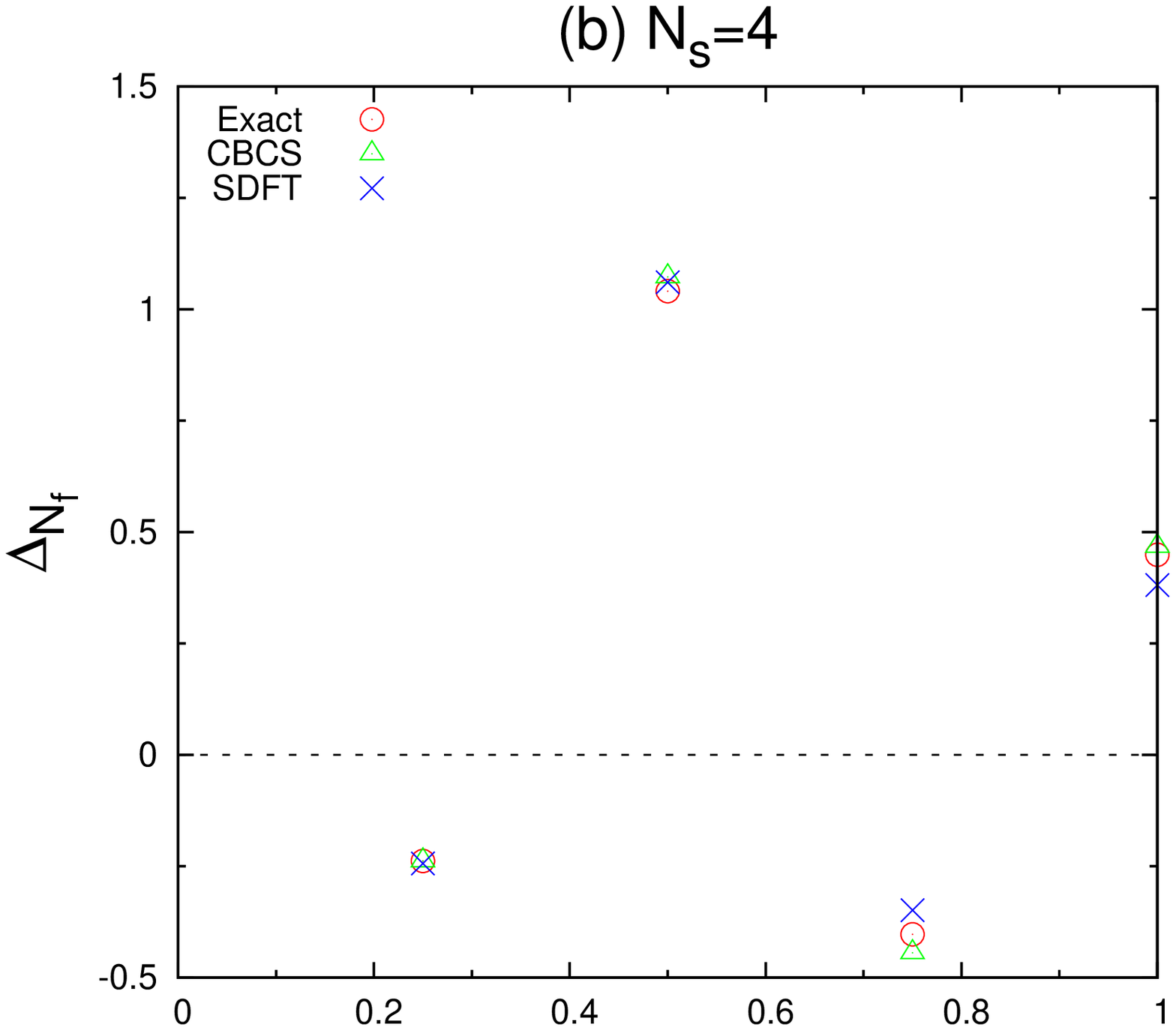}}&
\scalebox{0.5}[0.5]{\includegraphics{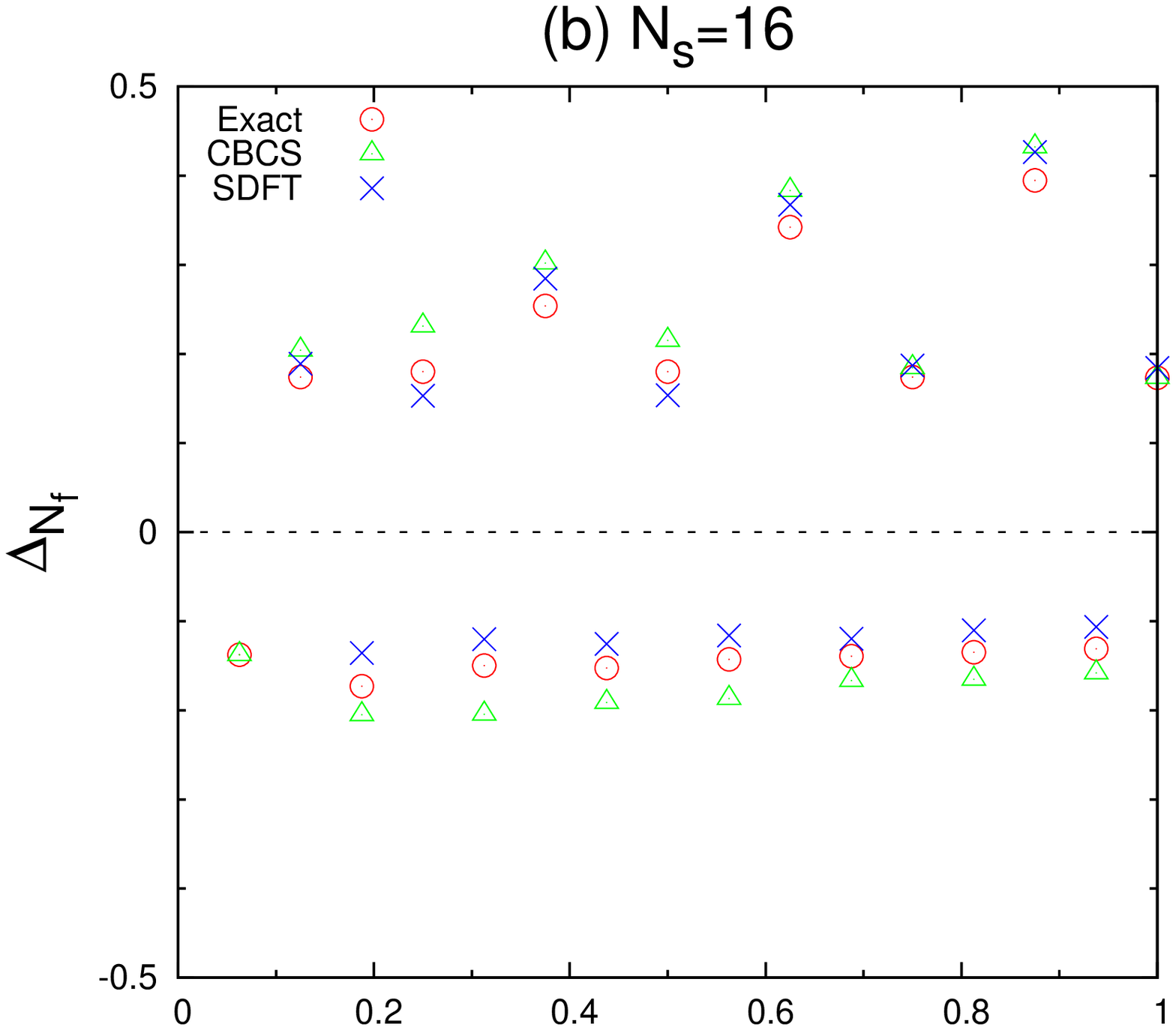}}\\
\scalebox{0.5}[0.5]{\includegraphics{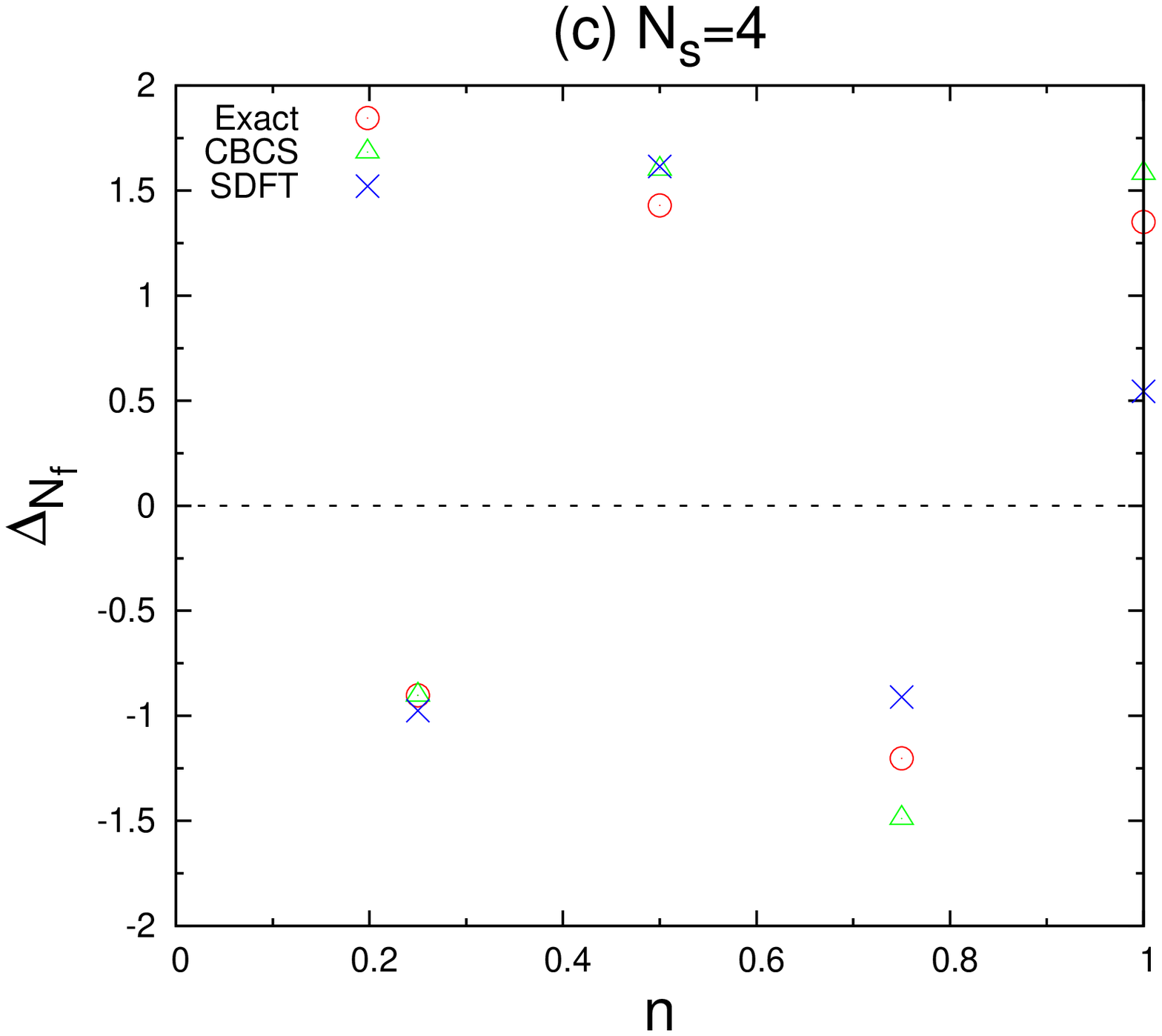}}&
\scalebox{0.5}[0.5]{\includegraphics{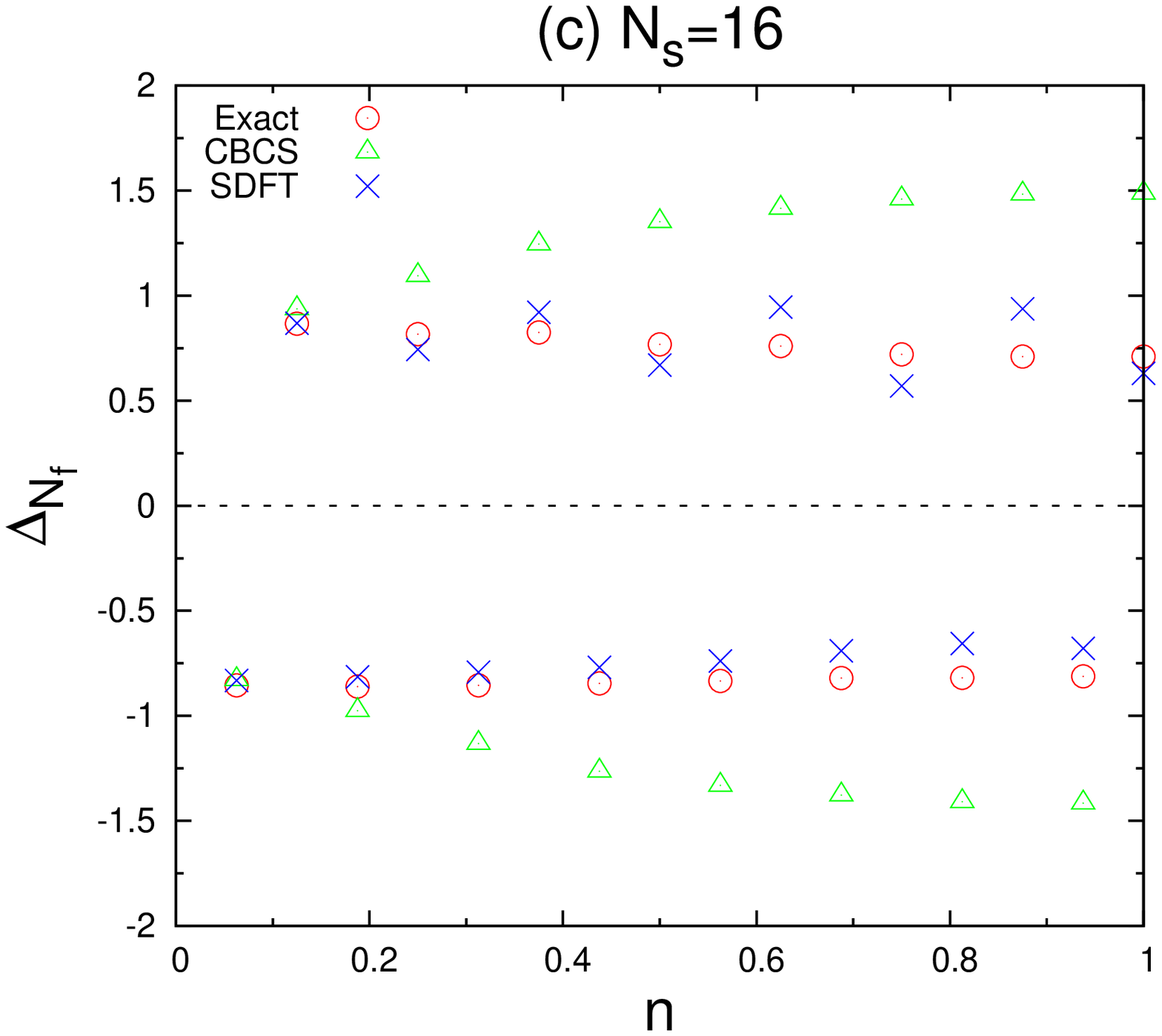}}
\end{tabular}
\caption{(Color online) The energy gap (in units of $t$) of the attractive 1D Hubbard
 model with periodic boundary condition as a function
 of the particle density $n = N_{\rm f}/N_{\rm s}$ is plotted for
 various coupling strength in a lattice of sites $N_{\rm s} = 4$ and
 $16$. The SDFT/BALSDA (crosses) results are compared with the exact
 (open circles) and CBCS data (upper triangle). The left panel is for
 $N_s=4$ and the right panel $N_{\rm s} = 16$. The gap is negative for
 odd $N_{\rm f}$ but positive for even $N_{\rm f}$ reflecting the
 pairing information induced by the negative interactions. (a) For
 $u=-1$; (b) For $u=-1.5$; and (c) For $u=-4$.
 For the system of even number of particles, the zero magnetization ($s=0$) is used.
 For the system of odd number of particles, we choose $N_\uparrow-N_\downarrow=1$, that is, $s=1/(2N_s)$.
 \label{fig:six}}
\end{center}
\end{figure*}

\begin{figure}
\begin{center}
\includegraphics*[width=1.0\linewidth]{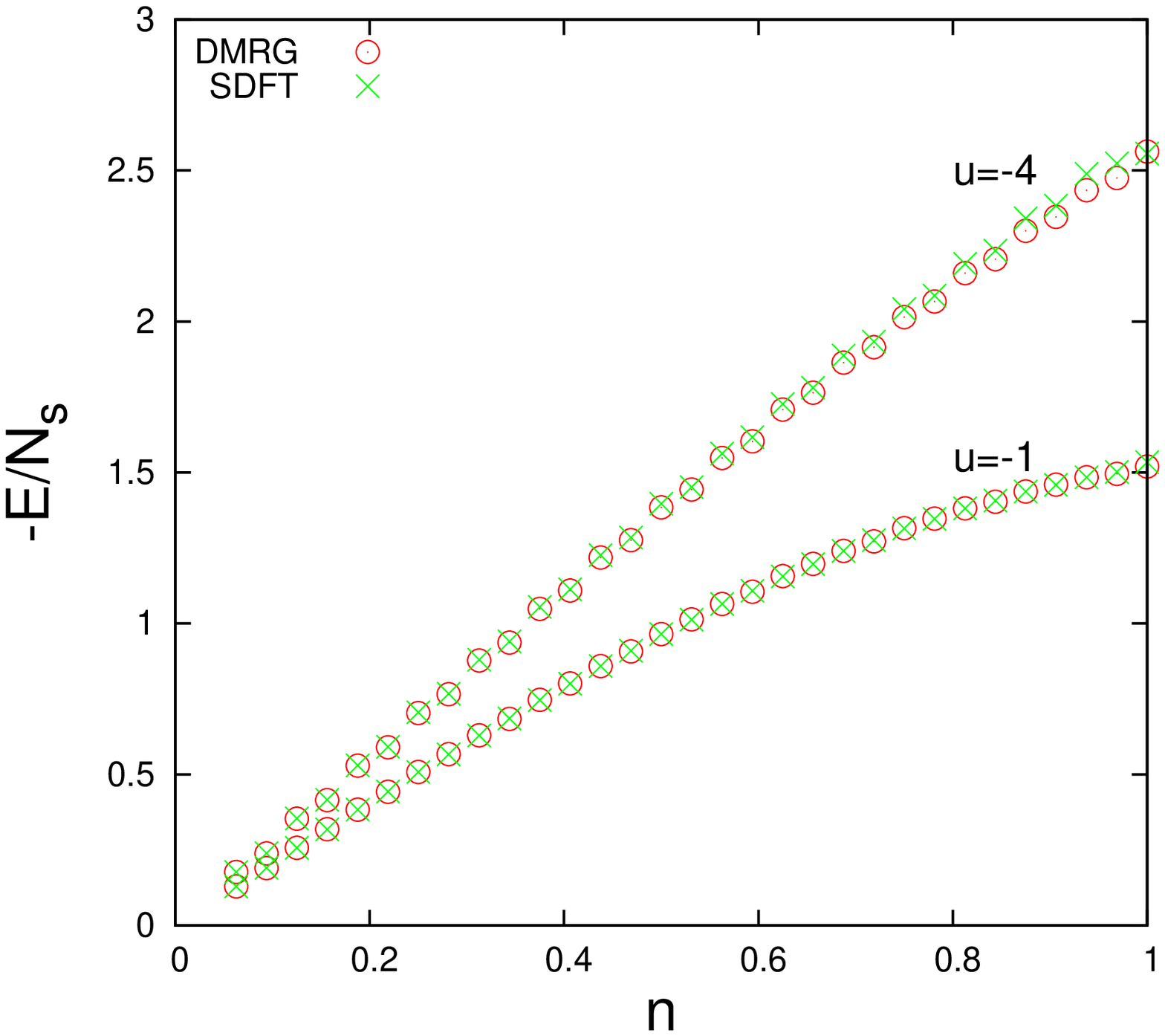}
\caption{(Color online) (a) The ground-state energy per site (in units of $t$) of the
 attractive 1D Hubbard model in a hard-wall as a
 function of particle density $n=N_{\rm f}/N_{\rm s}$ in a lattice of
 $N_{\rm s}=32$ sites for $u=-1$ and $-4$.\label{fig:seven}}
\end{center}
\end{figure}

\begin{figure}
\includegraphics*[width=1.00\linewidth]{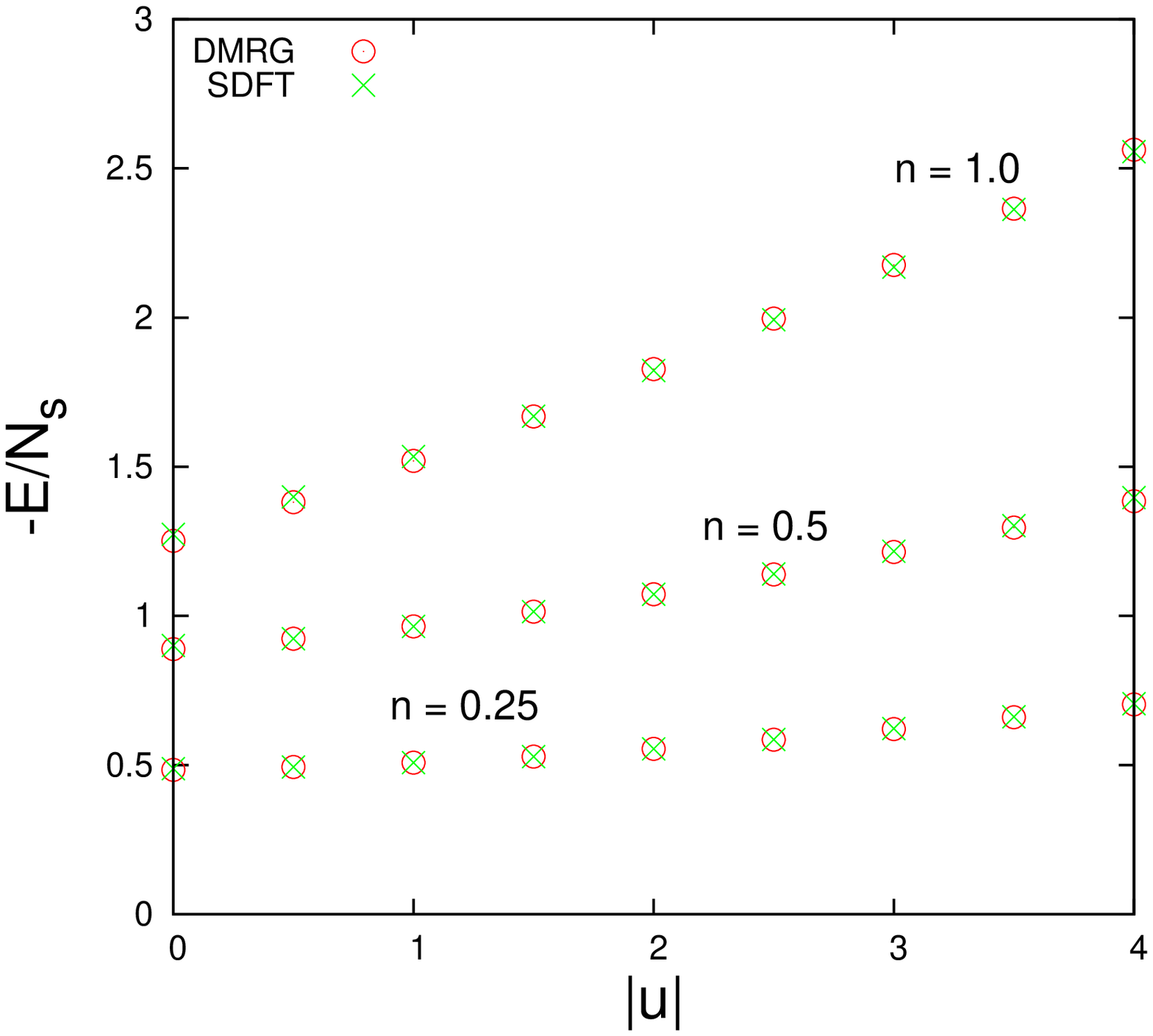}
\includegraphics*[width=1.00\linewidth]{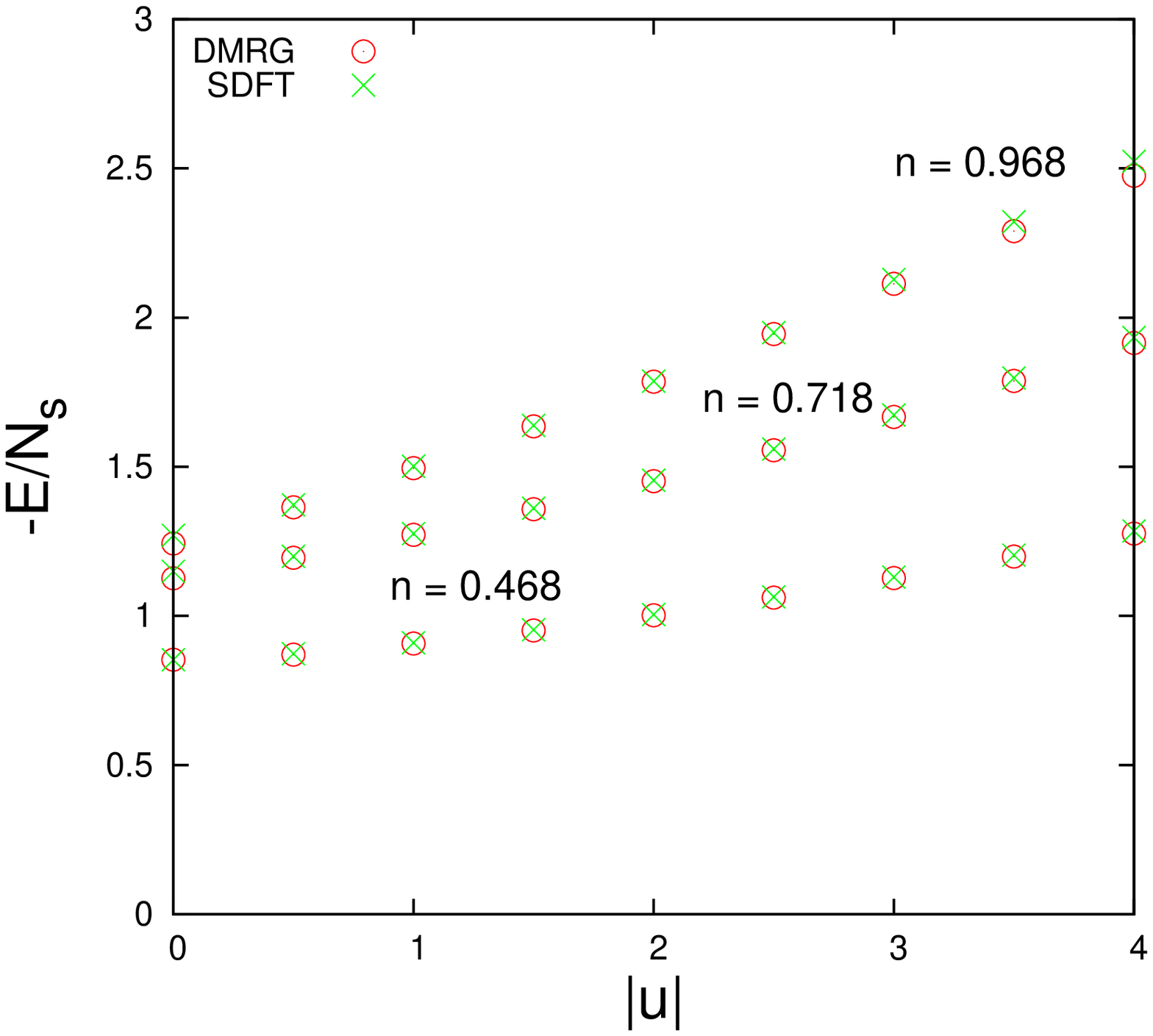}
\caption{(Color online) The ground-state energy per site (in units of $t$) of the
 attractive 1D Hubbard model in a hard-wall as a
 function of the coupling strength $u$ in a lattice of sites $N_{\rm
 s}=32$ for various values of fillings $n$.\label{fig:eight}}
\end{figure}

\begin{figure}
\centering
\includegraphics[width=1.0\linewidth]{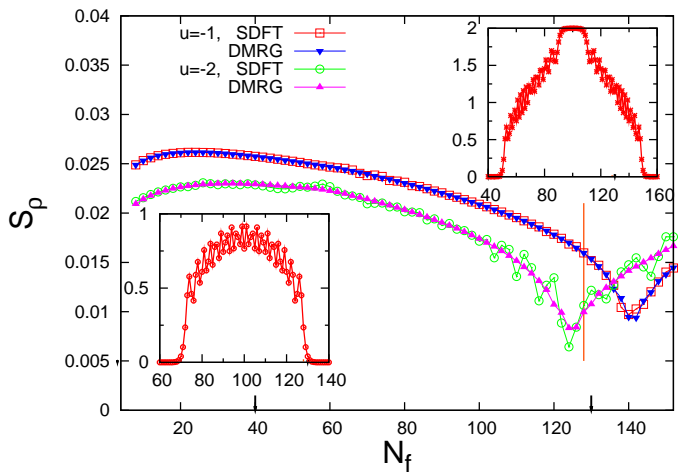}
\caption{(Color online) The thermodynamic stiffness $S_{\rho}$ (in units
 of $t$) as a function of $N_{\rm f}$ for $u=-1$ and $u=-2$ with $V_2/t=
 10^{-3}$, and $N_{\rm s}=200$ lattice sites. In the inset we show the
 density profile in the bottom (top) for $N_{\rm f}=40$ ($N_{\rm
 f}=132$) for $u=-2$. Their corresponding positions are indicated by the
 arrows in the $N_f$-axis. The vertical line indicates another possible
 quantum phase transition driven by the attractive interaction strength
 at fixed $N_{\rm f}=128$, which is checked through the concept of
 density-functional fidelity and fidelity susceptibility in
 Fig.~\ref{fig:ten}. \label{fig:nine}}
\end{figure}

\begin{figure}
\centering
\includegraphics*[width=1.0\linewidth]{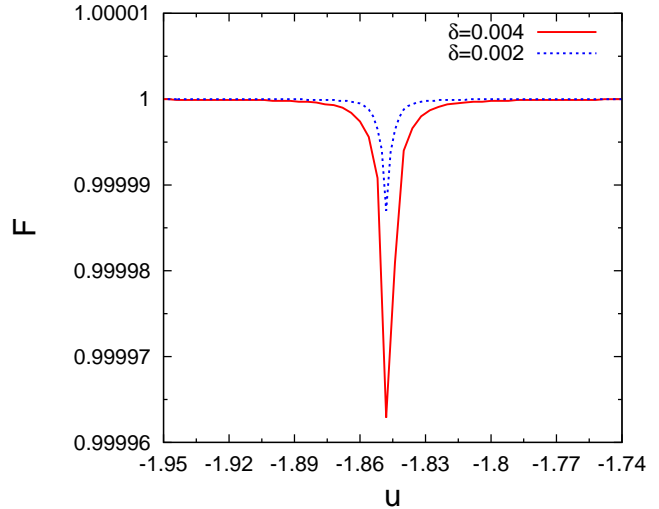}
\includegraphics*[width=1.0\linewidth]{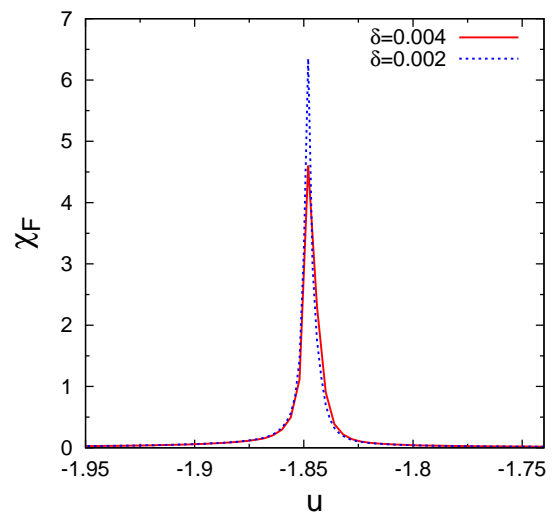}
\caption{(Color online) Density function fidelity (upper panel) and
 fidelity susceptibility (lower panel) as a function of couple strength
 $u$ for $N_{\rm s}=100$, $N_{\rm f}=128$, and $V_2/t =
 10^{-3}$. $\delta u=0.004$ (solid red line) and $\delta u=0.002$ (dotted blue line) are shown in the figures. \label{fig:ten}}
\end{figure}

In this section we present the numerical results obtained from the
self-consistent solution of Eqs.~(\ref{eq:sks})--(\ref{eq:closure})
based on the BALSDA of Eq.~(\ref{eq:balda}). We illustrate our main
numerical results, which are summarized in
Figs.~\ref{fig:four}--\ref{fig:ten}. We compare the results
obtained with the SDFT/BALSDA to results of DMRG and CBCS. In the use
of DMRG, the number of states kept is set maximum 1200 and the
truncation error is less than $10^{-10}$.


First, in Figs.~\ref{fig:four}--\ref{fig:six}, we show the results for
systems of periodic boundary condition. In Fig.~\ref{fig:four}, the GSE
per site (absolute value) is plotted as a function of the particle
density for $u=-2$ and $u=-4$. Good agreement is seen in the whole range
of fillings between the SDFT results and the exact Bethe-ansatz ones,
while CBCS calculations deviate in higher density. In
Fig.~\ref{fig:five}, we show the GSE as a function of interaction
strength for the system of even and odd number of particles. The
comparison shows that SDFT results give poorer agreement for small
system size with even number of particles than CBCS but better agreement
in other situations, such as for larger system size or stronger coupling
strength. We should keep in mind that the reasons for which the SDFT results deviate from the exact Bethe-ansatz ones are due to both finite size effects and
the xc functional built in the KS potential, while for the CBCS results they are affected by both finite-size effects and lack of correlation.
The ground-state energy and the energy gap (in Fig.~\ref{fig:six}, see below) are less accurate by SDFT in small size systems and by CBCS at intermediate coupling strengths.

For the 1D Hubbard model of attractive interaction, one defines the
gap or binding energy as,~\cite{marsiglio_prb1997,capellenew}
\begin{equation}\label{eq:gap}
\Delta_{N_{\rm f}} \equiv \frac{1}{2} \left[E_0 ({N_{\rm f} - 1}) - 2
E_0 ({N_{\rm f}}) + E_0({N_{\rm f} + 1}) \right] \,,
\end{equation}
which describes the energy difference between two systems, the one of
two subsystems with $N_{\rm f}$ particles each, and the other of $N_{\rm
f}+1$ and $N_{\rm f}-1$, respectively. From the viewpoint of DFT, this
gap has two contributions, the one from the KS gap, and the other from
the xc gap, related to the discontinuity of GSE as the number of
fermions varies across an integer.~\cite{LimaEPL} This definition is
also explained as the negative of the excitation gap, which equals to
the half of the energy it takes to break a pair in the system.~\cite{Blume}
In Fig.~\ref{fig:six}, the energy gap as a function
of the density is shown for (a) $u=-1$, (b) $u=-1.5$, and (c) $u=-4$ for
$N_{\rm s}=4$ (left panel), and $N_{\rm s}=16$ (right panel). The
comparison shows the SDFT calculations are qualitatively right in
describing the properties of the gap. Quantitatively we find that a
better agreement is achieved by SDFT for larger system size and stronger
attractive interaction. The gap is negative for odd number of particles,
and positive for even number of particles, which reflects the fact that
the system with negative interactions favors
pairing.~\cite{marsiglio_prb1997,Tanaka_prb1999,Emery} This is the
even-odd effects discussed by Tanaka and
Marsiglio.~\cite{Tanaka_prb1999,Amico} From the right panel of
Fig.~\ref{fig:six}, we notice that the energy gap oscillates as a
function of even particle numbers (depending on whether the number of
particles is $4m$ or $4m+2$), which is the super-even effect (for
details, see Ref.~\onlinecite{Tanaka_prb1999}).


The confined system composed of the hard wall, which can be realized in
experiments with the atom chips, focused laser beams and trapped
impurity atoms,~\cite{hard-wall-ex} breaks the inhomogeneity but still
can be solved through the Bethe-ansatz method.~\cite{hardwall} In
Figs.~\ref{fig:seven} and \ref{fig:eight}, we show the SDFT results for
the GSE compared to DMRG ones. The system is of $N_{\rm s}=32$ lattice
sites with $u=-1$ and $-4$. We find that, in almost all the range of particle
density and weak-to-intermediate interaction strength, SDFT results give
an excellent agreement with DMRG calculations. Small deviations happen
around the half filling.

Next, we discuss a 1D system of Fermi gases trapped in a harmonic
potential, which is a system that breaks the homogeneity and the
integrability of the model and can not be exactly solved. A
quasi-classical approach called the local density approximation is
usually used to study the density distribution profile when the external
potential is slowly varying.~\cite{LDA} Here and in the following the
DFT based on LDA is used. We would like to mention that for the system of even
number of particles and spin-unpolarized, the DFT is used. But for the
system of odd number of particles, the SDFT has to be used. Our results
are shown in Table \ref{table:one} and \ref{table:two} and
Figs.~\ref{fig:nine} and \ref{fig:ten}. In Table~\ref{table:one}, we
show the GSE for small systems of lattice size $N_{\rm s}=100$ and large
systems of $N_{\rm s}=300$ for the weak-to-intermediate interaction
strength. The corresponding pair-binding energies are calculated
accordingly, which are defined as the energy difference between two
systems, the one with two subsystems of particles $N_{\rm f}$ and
$N_{\rm f}+2$, the other with two subsystems of $N_{\rm f}$ particles
each,
\begin{equation}\label{bindE}
{\cal E}_{\scriptscriptstyle P} (N_{\rm f}) = E_0 (N_{\rm f} + 2) + E_0
 (N_{\rm f}) - 2 E_0 (N_{\rm f} + 1)\,.
\end{equation}
Our calculation shows that the pair-binding energies are negative, which
illustrates that the system likes to accommodate pairs and has the
tendency to form stable dimers. Furthermore, we calculate the
four-particle binding energies,~\cite{Schmidt}
\begin{equation}
{\cal E}_4 (N_{\rm f})= \Delta E_4 (N_{\rm f})-2 \Delta E_2(N_{\rm f})
 \,,
\end{equation}
where $\Delta E_i (N_{\rm f})$ is the energy difference defined as
$\Delta E_i (N_{\rm f}) = E_0 (N_{\rm f} + i) - E_0 (N_{\rm f})$.
A bound state of two particles is confirmed by observing that, in
Table~\ref{table:two}, ${\cal E}_p (N_{\rm f}) < 0$ and ${\cal
E}_4(N_{\rm f}) >0$, which shows the evidence of a paired superfluid. We
conclude that, the system induces effective attractions between
particles leading to pairs formation (${\cal E}_p(N_{\rm f}) <0$) and
effective repulsions between pairs (${\cal E}_4 (N_{\rm f}) >0$)
avoiding clusters formation. The same calculation is done for the 1D
Fermi-Hubbard model with repulsive interactions. It is found that ${\cal
E}_p (N_{\rm f}) >0$ and ${\cal E}_4 (N_{\rm f}) >0$ and no paired
superfluid is formed. The comparison of the GSE between SDFT and DMRG
shows that the agreement is up to the second decimal digit, except for
few cases, such as for $N_{\rm f}=30$ and $u=-2$ with the relative error
of 1.2\%.

\begin{table}[htbp]
\centering
\caption{Ground-state energies per site (in units of $t$) with harmonic
 trap $V_{\rm ext}/t=4 \times 10^{-3}$, and $V_{\rm ext}/t=1 \times 10^{-4}$
 for the system of $N_{\rm s}=100$, $N_{\rm f}=30$ and $N_{\rm s}=300$,
 $N_{\rm f}=90$ respectively. The couple strength are $u=-0.5$, $-1$,
 $-1.5$, and $-2$. The SDFT results are compared to the DMRG
 ones. $|{\Delta E}_0|$ is the relative error between the SDFT and DMRG
 calculations.
 \label{table:one}}.
\begin{center}
\begin{tabular}{ccccccccc}
\hline \hline $u$ & $V_{\rm ext}/t$ & $N_{\rm S}$ & $N_{\rm f}$ &
 $E_0$(SDFT)&   $E_0$(DMRG) &  $|{\Delta E}_0|$\\
$-0.5$  & $ 4\times 10^{-3}$  & $100$  & $30$ & $-0.358674$ &$-0.358318$
&$0.1$\\
& & & $31$ & $-0.361838$ &$-0.361488$ &$0.1$ & \\
& & & $32$ & $-0.365213$ &$-0.364864$ & $0.1$& \\
& & & $34$ & $-0.369698$ &$-0.369352$ & $0.1$& \\
& $10^{-4}$ & $300$ & $90$ &$-0.489727$ &$-0.489200$ &$0.1$ \\
& & & $91$ & $-0.493653$ &$-0.493265$ & $0.1$& \\
& & & $92$ & $-0.497871$ &$-0.497353$ & $0.1$& \\ \hline
$-1$ & $4\times 10^{-3}$ & $100$ & $30$ & $-0.394805$ &$-0.393394$
&$0.4$ \\
& & & $31$ & $-0.399480$ &$-0.398080$ &$0.4$ & \\
& & & $32$ & $-0.404714$ &$-0.403324$ &$0.3$ & \\
& & & $34$ & $-0.412661$ &$-0.411302$ &$0.3$ & \\
& $10^{-4}$ & $300$ & $90$ & $-0.516192$ &$-0.515279$ &$0.2$ \\
& & & $91$ & $-0.520473$ &$-0.519706$ &$0.1$ & \\
& & & $92$ & $-0.525122$ &$-0.524224$ & $0.2$& \\ \hline
$-1.5$ & $4\times 10^{-3}$ & $100$ & $30$ & $-0.435840$
&$-0.432676$&$0.7$ \\
& & & $31$ & $-0.441930$ &$-0.438820$ &$0.7$ & \\
& & & $32$ & $-0.449298$ &$-0.446184$ &$0.7$ & \\
& & & $34$ & $-0.460821$ &$-0.457841$ &$0.7$ & \\
& $10^{-4}$ & $300$ &$90$ & $-0.547965$ &$-0.546426$ &$0.3$ \\
& & & $91$ & $-0.552561$ &$-0.551187$ &$0.2$ & \\
& & & $92$ & $-0.557752$ &$-0.556250$ &$0.3$ & \\ \hline
$-2$ & $4\times 10^{-3}$ & $100$ & $30$ & $-0.481959$ &$-0.476300$
&$1.2$ \\
& & & $31$ & $-0.489022$ &$-0.483732$ &$1.1$ & \\
& & & $32$ & $-0.498967$ &$-0.493580$ &$1.1$ & \\
& & & $34$ & $-0.514133$ &$-0.509107$ &$1.0$ & \\
& $10^{-4}$ & $300$ &$90$ & $-0.584805$ &$-0.582994$ &$0.3$ \\
& & & $91$ & $-0.589692$ &$-0.588015$ &$0.3$ & \\
& & & $92$ & $-0.595528$ &$-0.593786$ &$0.3$ & \\ \hline \hline
\end{tabular}
\end{center}
\end{table}
\begin{table*}[htbp]
\centering
\caption{Pair- and four-particle binding energies ${\mathcal E}_p$ and
 ${\mathcal E}_4$ per site (in units of $t$) are calculated with the
 energies from Table \ref{table:one} for the system of
 $N_{\rm f}=30$ and $N_{\rm s}=100$ under the presence of a harmonic trap
 $V_{\rm ext}/t=4 \times 10^{-3}$. The couple strength are $u=-0.5$, $-1$,
 $-1.5$, and $-2$. The SDFT results are compared to the DMRG
 ones. $\Delta{\mathcal E}_{}$ is the relative error between the SDFT
 and DMRG calculations.
 \label{table:two}}.
\begin{center}
\begin{tabular}{cccccccc}
\hline \hline
 $u$ & ${\mathcal E}_{p}$ (SDFT) & ${\mathcal E}_{p}$ (DMRG)&
 $\Delta{\mathcal E}_{p}$ & ${\mathcal E}_{4}$ (SDFT) & ${\mathcal
 E}_{4}$ (DMRG) & $\Delta{\mathcal E}_{4}$&\\
\hline
 -0.5  &-0.000211  &-0.000207 &1.9&0.002054 &0.002058  & 0.2& \\ \hline
 -1  &-0.000559  &-0.000559 &0.1 &0.001962 &0.001952   & 0.5& \\  \hline
 -1.5  &-0.001278  &-0.001219 &4.8&0.001935 &0.001850 & 4.6&\\ \hline
 -2  &-0.002881  &-0.002417 &19.2&0.001842 &0.001754   & 5.0& \\ \hline
 \hline
\end{tabular}
\end{center}
\end{table*}

The harmonically trapped system described by Eq.~(\ref{eq:hubbard})
allows for two different spatial coexistent quantum phases. The one is
the phase characterized by the Luther-Emery (LE) liquid combining with the
appearance of the atomic density wave ($n_i<2$), and another is
characterized by a composite phase with a plateau in the trap center of
tightly bound spin-singlet dimers ($n_i=2$) surrounded by Luther-Emery
layers ($n_i<2$). We call the latter the band-insulating (BI) phase.
These two different phases can be well classified by a
scaled dimensionless variable, the characteristic density
$\tilde{\rho}=N_{\rm f} \sqrt{V_2 / t}$ and the interaction strength $u$
in the $\tilde{\rho}$--$u$ phase diagram~\cite{Rigol2004}. Thus for
certain $u$, the quantum phases in the presence of confinement can be
driven from one to another by increasing $\tilde{\rho}$. Here we study
this phenomena by investigating the thermodynamic stiffness, which is
defined as the inverse of the global compressibility,
\[
S_\rho=\frac{\delta \mu}{\delta {N_{\rm f}}}=\frac{\delta^2 E_0}{\delta
{N^2_{\rm f}}}\,,
\]
where $\mu$ is the global chemical potential. The stiffness has been
successfully used in the confined system to characterize the quantum
phase transitions. The different coexisted phases appeared in the
harmonically confined system are well separated by the nonanalyticity point
in the stiffness. The increase of stiffness is related to the
incompressible nature of the insulating phases.~\cite{gaoprb73rapid} For
the system we studied, the phase is of Luther-Emery type for small
$\tilde{\rho}$~\cite{Hu} and a composite phase of Luther-Emery-liquid
type in the wings and insulting type in the bulk for large
$\tilde{\rho}$. We test for a system in a confinement of strength
$V_2/t=10^{-3}$ and $N_{\rm s}=200$ lattice sites. The comparison of
SDFT with DMRG shows that, in Fig.~\ref{fig:nine}, the stiffness calculated
from SDFT is quantitatively correct for weak interaction strength of
$u=-1$ and qualitatively correct for $u=-2$. The large deviation happens
for the system of large number of particles. We find by DMRG, the phase transition
happens at $N_{\rm f}=141$  ($N_{\rm f}=125$) for $u=-1$ ($u=-2$),
driven by the increasing number of the particles, while the results from
SDFT slightly deviate from the DMRG ones with $N_{\rm f}=140$  ($N_{\rm
f}=124$) for $u=-1$ ($u=-2$). We notice that the bulk compressibility or
the stiffness calculated here is not a real singular quantity since it
includes the sizable contributions from the liquid wings of the density
profile.~\cite{Leo}

Finally we study the phase transition driven by the attractive
interaction strength through the density-functional fidelity, which is
defined as \cite{fidelity_Gu}
\begin{equation}
F(u, u^\prime )=\frac{1}{N_f}\text{tr}\sqrt{n_i(u )n_i(u^{\prime })}\,,
\end{equation}
here $n_i(u)$ is the density of particles in lattice site $i$ with
interaction strength $u$. If we fix the distance between $u$ and
$u^{\prime}$, $\delta u=u-u'$, the density-functional fidelity is
expected to show a drop around the critical point due to the maximum
distance between the two ground states in different quantum
phases.~\cite{fidelity_Gu} In the top panel of Fig.~\ref{fig:ten}, the
density-functional fidelity is plotted as a function of the coupling
strength $u$ with the distance between $u$ and $u^{\prime}$ chosen as $0.002$ and $0.004$. We notice
that the fidelity drops abruptly at the critical point $u=-1.848$,
which reflects the fact that the two ground states in the different quantum
phases have the maximum distance. In other words, when we fix the number
of fermions and the amplitude of the external potential by varying $u$,
this critical value of the attractive interaction is obtained. At that point, the system is in
the BI phase with the localized site occupancy of $n_i=2$ in the center of the trap.

The fidelity susceptibility, the leading term of the fidelity, is found
to play a central role in the fidelity approach in precisely
characterizing the quantum phase transition, which is defined as,
\begin{equation}\label{eq:FS}
F(u, u+\delta u)=1-\frac{(\delta u)^2}{2}\chi_{\rm F}\,,
\end{equation}
where the fidelity susceptibility (FS) $\chi_{\rm F}$ takes the form
\[
\chi_{\rm F} = \sum_i \frac{1}{4n_i} \left( \frac{\partial n_i}{\partial
u} \right)^2 \,.
\]
In the bottom panel of Fig.~\ref{fig:ten}, the FS is shown as a function
of the coupling strength $u$, which manifests itself as a marked peak at
the same critical point as that in the fidelity function. The sharp peak
in the FS implies principally that the density distribution evolves
dramatically around the critical point, which is therefore taken as
another important quantity to detect the quantum phase
transition.~\cite{fidelity_Gu,Zanardi06} The peak in the FS is free from
the arbitrariness of the small parameter $\delta u$. To prove
this, we calculated the FS with two different $\delta u$ in
Fig.~\ref{fig:ten}(b). Both cases have good agreements and support the
conclusion that the fidelity susceptibility does not depend on the value
of $\delta u$, but the fidelity does.~\cite{You}

\section{summary}

In summary, in this paper we have studied the ground-state properties of
two-component Fermi gases of attractive interactions in a 1D system, by
performing SDFT and DMRG simulations. Three different systems have been
tested, for fermions on a lattice line with periodic boundary condition,
in a hard wall or in the presence of a harmonic confinement. For the
system of periodic boundary condition or hard-wall, we showed the
calculations for the GSE and the energy gap in accordance with the exact
Bethe-ansatz, CBCS, and DMRG results. We found that the ground-state
energy is faithfully reproduced by SDFT in the large range of
interaction strength and fillings. For the system under the harmonic
confinement, we calculated the pair-binding and
four-particle binding energies to show that the system of attractive interactions induces effective
attractions between particles leading to pairs formation and effective repulsions between pairs avoiding clusters formation.
The system is either in the Luther-Emery type phase or in the
composite phase of Luther-Emery-like in the wings and insulating like in
the bulk. In order to characterize the quantum phase transition driven by the
particle number or the attractive interaction strength, the thermodynamic stiffness, the
density-functional fidelity, and the fidelity susceptibility are studied, respectively.
We found that the thermodynamic stiffness is able to locate the quantum critical point driven by
the particle number through the nonanalyticity point, and $F$ ($\chi_F$) is able to locate the quantum critical point driven by the
attractive interaction through the abrupt drop (the sharp peak) at the critical point.

\begin{acknowledgments}
This work was supported by NSF of China under Grant Nos.~10974181 and
10704066, the Program for Innovative Research Team at Zhejiang Normal University, a Grant-in-Aid
for Scientific Research (Grant No.~20500044) and Priority Area ``Physics
of New Quantum Phases in Superclean Materials'' (Grant No.~20029019)
from the Ministry of Education, Culture, Sports, Science and
Technology, Japan. G.X. gratefully acknowledges many useful discussions
with Shi-Jian Gu and Shu Chen. The authors are grateful to Frank
Marsiglio for making the data of numerical calculations of
Ref.~\onlinecite{Tanaka_prb1999} available to us. DMRG results are
checked in part using ALPS DMRG application.~\cite{ALPS}
\end{acknowledgments}

\end{document}